\documentclass[12pt]{article}

\usepackage{amssymb}
\usepackage{amsfonts}
\usepackage{caption}
\usepackage{eurosym}
\usepackage{geometry}
\usepackage{ulem}
\usepackage{graphicx}
\usepackage{caption}
\usepackage{color}
\usepackage{setspace}
\usepackage{sectsty}
\usepackage{comment}
\usepackage{footmisc}
\usepackage{caption}
\usepackage{pdflscape}
\usepackage{subfigure}
\usepackage{array}
\usepackage{hyperref}
\usepackage{soul}
\usepackage{caption}
\usepackage{array}
\usepackage{anyfontsize}
\usepackage{authblk}

\newcolumntype{L}[1]{>{\raggedright\let\newline\\arraybackslash\hspace{0pt}}m{#1}}
\newcolumntype{C}[1]{>{\centering\let\newline\\arraybackslash\hspace{0pt}}m{#1}}
\newcolumntype{R}[1]{>{\raggedleft\let\newline\\arraybackslash\hspace{0pt}}m{#1}}
\usepackage[
backend=biber,
style=authoryear-ibid,
sorting=nyt,
uniquename=false
]{biblatex}
\title{References: \texttt{biblatex} package}
\citetrackerfalse

\usepackage[font=normalsize,labelfont={bf}]{caption}

\begin{document}
\begin{titlepage}
\title{Colocation of skill related suppliers -- Revisiting coagglomeration using firm-to-firm network data}


\author[1,2,*]{Sándor Juhász}
\author[2,3]{Zoltán Elekes}
\author[2,4]{Virág Ilyés}
\author[1]{Frank Neffke}

\renewcommand\Affilfont{\footnotesize}
\renewcommand\Authsep{, }
\renewcommand\Authands{, }
\renewcommand\Affilfont{\footnotesize}

\affil[1]{\footnotesize Complexity Science Hub, Vienna, Austria}
\affil[2]{\footnotesize ANETI Lab, HUN-REN Centre for Economic and Regional Studies, Budapest, Hungary}
\affil[3]{\footnotesize CERUM, Umeå University, Umeå, Sweden}
\affil[4]{\footnotesize ANETI Lab, Corvinus University of Budapest, Budapest, Hungary}
\affil[*]{\footnotesize Corresponding author: \href{mailto:juhasz@csh.ac.at}{juhasz@csh.ac.at}}

\date{}

\maketitle
\begin{abstract}
\begin{singlespace}
\noindent Strong local clusters help firms compete on global markets. One explanation for this is that firms benefit from locating close to their suppliers and customers. However, the emergence of global supply chains shows that physical proximity is not necessarily a prerequisite to successfully manage customer-supplier relations anymore. This raises the question when firms need to colocate in value chains and when they can coordinate over longer distances. We hypothesize that one important aspect is the extent to which supply chain partners exchange not just goods but also know-how. To test this, we build on an expanding literature that studies the drivers of industrial coagglomeration to analyze when supply chain connections lead firms to colocation. We exploit detailed micro-data for the Hungarian economy between 2015 and 2017, linking firm registries, employer-employee matched data and firm-to-firm transaction data from value-added tax records. This allows us to observe colocation, labor flows and value chain connections at the level of firms, as well as construct aggregated coagglomeration patterns, skill relatedness and input-output connections between pairs of industries. We show that supply chains are more likely to support coagglomeration when the industries involved are also skill related. That is, input-output and labor market channels reinforce each other, but supplier connections only matter for colocation when industries have similar labor requirements, suggesting that they employ similar types of know-how. We corroborate this finding by analyzing the interactions between firms, showing that supplier relations are more geographically constrained between companies that operate in skill related industries.
\\
\end{singlespace}

\vspace{0in}
\vspace{0in}

\bigskip
\end{abstract}
\setcounter{page}{0}
\thispagestyle{empty}
\end{titlepage}
\pagebreak \newpage


\onehalfspacing

\section{Introduction} \label{sec:introduction}

Fuelled by a dramatic decrease in transportation costs, global value chains (GVCs) nowadays span across countries and continents  \parencite{crescenzi2023harnessing, baldwin2022gvc, johnson2018gvc}. Yet, buyer-supplier transactions still surprisingly often take place over short distances \parencite{bernard2019jpe}. We argue that this puzzle can be in part resolved by focusing on the heterogeneity in existing value chain linkages: not all customer-supplier relations will be equally amenable to long-distance interaction. In particular, we propose that existing spatial patterns of value chain connections reflect the degree to which value chain partners use similar know-how, which would allow them to embed significant amounts of tacit knowledge in their transactions. We evaluate this conjecture by studying the coagglomeration of industries in Hungary, a country that is deeply embedded in transnational value chains and for which rich micro-data exist that describe firms' workforces as well as their transactions with other firms.

Following \textcite{egk2010coagglom}, a growing literature has analyzed agglomeration externalities by studying coagglomeration \parencite{helsley2014coagglom, faggio2017heterogagglom, howard2013vietnam, gabe2016occupcoagg, gallagher2013indcoagg, bertinelli2005belgium, aleksandrova2020russcoagg, kolko2010coaggservices}. One strand of this literature has focused on differences between various sectors in their propensity to coagglomerate, highlighting, for instance, differences between manufacturing and service industries \parencite{diodato2018coagglom, oclery2021unravelling}. So far, these studies have focused on the \emph{activities} that coagglomerate, neglecting heterogeneity in the \emph{links} between them. More importantly, the three Marshallian agglomeration forces -- labor pooling, value chain linkages and knowledge spillovers -- have so far been treated as if they acted independently from one another. In contrast, we argue that these forces can reinforce one another.

We propose that, while firms coagglomerate to facilitate labor pooling and buyer-supplier relations, these two channels do not operate independently. Instead, we expect that value chain partners are more likely to coagglomerate if they operate in skill related \parencite{neffke2013skillrel} industries. This proposition is based on the idea that much crucial knowledge underpinning a firm's competitive advantage resides in its human capital \parencite{kogut1992knowledge, spender1996knowledge}. Firms that require workers with similar skills are cognitively proximate and thus likely to be able to share knowledge and learn from one another \parencite{neffke2013skillrel, neffke2017interind}. Moreover, some intermediate products require a clear understanding of how they are produced to effectively handle them. In such cases, buyers may need to increase the cognitive proximity to their suppliers. Because geographical proximity facilitates both interfirm learning and coordination involving the exchange of tacit knowledge \parencite{jaffe1993spillovers, audretsch1996spillovers}, value chain interactions will benefit most from colocation if the industries involved are also skill related.

We provide empirical support for this hypothesis, using uniquely detailed administrative datasets from Hungarian public registers. These data cover all companies operating in Hungary, 50\% of their employees, as well as value-added tax records for buyer-supplier transactions among them. Building on the literature on coagglomeration, skill relatedness and production networks, we use these data to construct measures of coagglomeration, skill relatedness and input-output connections between detailed industries. 

We use these data first to show that \textcite{egk2010coagglom}'s finding that value chain and labor pooling links drive coagglomeration in the US can be replicated for Hungary. To do so, we rely on the same instrumental variables approach, instrumenting skill relatedness and input-output linkages in Hungary with analogous quantities calculated in other countries. Next, we show that input-output connections between industries only lead to substantial coagglomeration if industries are also skill related. In contrast, skill related industries always display substantial coagglomeration tendencies, regardless of whether or not they are connected in value chains. Finally, we corroborate these aggregate findings at the micro-level by studying detailed spatial patterns of interfirm ties. 

Our study contributes to several strands of the literature. First, it adds to an expanding literature on industrial coagglomeration \parencite{egk2010coagglom, helsley2014coagglom, delgado2015defining, faggio2017heterogagglom, diodato2018coagglom, steijn2022dynamics}. Our main contribution to this literature is that we uncover important interactions between coagglomeration forces. Furthermore, we add evidence from Hungary, a small open economy where most inputs need to be imported \parencite{halpern_et_al2015aer}. The exceptionally rich data for the Hungarian economy allow us to go beyond studying \emph{potential} linkages between industries in terms of skill relatedness and input-output coefficients to \emph{actual}, observed, interactions between firms, strengthening the micro-foundations in this area of research. Second, our conceptual framework connects the literature on coagglomeration to discussions on buyer-supplier linkages in regional innovation systems \parencite{cooke_morgan1994ijtm, cooke1996sbe} and in GVCs \parencite{crescenzi2023harnessing, baldwin2022gvc, johnson2018gvc, boschma2024evolutionary}. Third, our study relates to the field of economic complexity analysis \parencite{hidalgo_hausmann2009pnas, hidalgo2021review, balland2022paradigm}. In particular, many product and industry spaces that are used in this literature are derived from coagglomeration patterns. By analyzing the drivers of coagglomeration patterns, we therefore also shed light on the forces that are captured in the product and industry spaces of economic complexity analysis. 

The paper is structured as follows. \textit{Section~\ref{sec:literature}} reviews prior literature and derives hypotheses. \textit{Section~\ref{sec:setting}} describes the data and construction of coagglomeration, skill relatedness and input-output metrics. \textit{Section~\ref{sec:result}} reports our empirical findings.  \textit{Section~\ref{sec:conclusion}} concludes with a discussion of implications, limitations and open questions.

\section{Colocation of industries and firms} \label{sec:literature}

\subsection{Coagglomeration} \label{sec:subsec_coagglomeration}

A core insight in economic geography is that firms seek each other's proximity to benefit from so-called agglomeration externalities. Accordingly, knowledge spillovers and access to pools of specialized labor and suppliers provide strong rationales for firms to colocate \parencite{marshall1920principles}. This results in an economic landscape  characterized by marked spatial clusters of related industries \parencite{delgado2014clusters}. \textcite{marshall1920principles}'s original account of why such industrial districts form pointed to access to specialized suppliers, skilled labor and knowledge: firms choose to colocate with their competitors because accessing these resources becomes harder as distances increase. In spite of substantial decreases in the cost of transporting goods, people and ideas \parencite{glaeser_kohlhase2004pirs, agrawal_goldfarb2008aer, catalini_et_al2020ms}, geographical clusters of firms are still thought to be core drivers of firms' competitiveness \parencite{porter_1990book, porter1998hbr}.

In parallel, a large literature in economic geography and urban economics has studied the role of agglomeration externalities in the success of local industries \parencite{glaeser_et_al1992jpe, henderson_et_al1995jpe, rosenthal_strange2004chapter, beaudry_schiffauerova2009rp, caragliu_et_al2016eg}. However, disentangling the relative importance of different agglomeration forces proved hard, because they act concurrently and all three forces lead to the same observable outcome: firms in the same industry will concentrate geographically. A breakthrough was achieved by \textcite{egk2010coagglom} who focused, not on agglomeration patterns  of individual industries, but instead on coagglomeration patterns of \emph{pairs} of industries that differ in the degree to which they share supplier, skill or knowledge relations. Doing so allowed the authors to show that coagglomeration is best explained by input-output dependencies, but that the other two forces, labor pooling and technological spillovers, play significant roles as well. 

The work of \textcite{egk2010coagglom} has sparked an expanding literature on coagglomeration \parencite{helsley2014coagglom, faggio2017heterogagglom, howard2013vietnam, mukim2015india, gabe2016occupcoagg, gallagher2013indcoagg, bertinelli2005belgium, aleksandrova2020russcoagg, kolko2010coaggservices}. One strand of this literature has focused on heterogeneity in coagglomeration forces. For instance, focusing on heterogeneity across time, \textcite{diodato2018coagglom} show that, over the course of the 20th century, the relative importance of labor pooling has increased to the point that it has surpassed input-output linkages as an explanation for coagglomeration patterns. \textcite{steijn2022dynamics} found a similar decrease in the relative importance of input-output linkages over time. However, these authors also highlight the increasing role of knowledge spillovers, identified through data on technological relatedness between industries, which surpassed even that of labor pooling. Moreover, they show that these shifts are, at least in part, brought about by increasing import penetration, a decrease in transportation costs and a fall in routine tasks. 

The relative importance of different Marshallian channels does not only change over time, but also varies across economic activities. For instance, \textcite{diodato2018coagglom} show that coagglomeration of service industries tends be more sensitive to labor sharing opportunities than coagglmeration of manufacturing industries. Whereas existing literature has so far studied how agglomeration forces vary across time and across economic activities, our analysis focuses on heterogeneity in the forces themselves. As we will argue below, agglomeration forces are unlikely to act in isolation, but rather may reinforce each other.

\subsection{Relatedness and economic complexity analysis} \label{sec:subsec_relatedness_eci}

Another body of related work is the literature on economic complexity analysis \parencite{hidalgo2007product, hidalgo_hausmann2009pnas} and in particular its adoption in evolutionary economic geography (EEG) \parencite{boschma2015uscities, balland_rigby2017eg, balland2022paradigm, mewes_broekel2022rp}. This literature argues that places develop by accumulating complementary capabilities that together allow economies to engage in complex economic activities \parencite{hidalgo_hausmann2009pnas, hidalgo2015book, frenken_et_al2023cjres}. Because many capabilities are hard to access from outside the region \parencite{neffke2018agents, frenken_et_al2023cjres}, economic development often takes the shape of a branching process in which economies expand by diversifying into activities that are closely related to their current activities \parencite{frenken_boschma2007jeg, neffke_et_al2011eg, hidalgo2018princple}.

Empirical work in economic complexity analysis often constructs abstract spaces that connect industries that are ``related''. These so-called industry spaces are relevant to our analysis in two ways. First, the most widely used industry (or product) spaces are, in fact, based on coagglomeration patterns \parencite{hidalgo2007product, hidalgo2021review, li2023evaluating}. Consequently, there is a direct link between economic complexity analysis and the coagglomeration literature. In this light, the coagglomeration literature can be seen as an effort to understand the underlying factors that are captured in prominent industry spaces.

Second, economic complexity analysis has constructed industry spaces using information other than coagglomeration patterns. A particularly relevant industry space is based on \textcite{neffke2013skillrel}'s ``revealed skill relatedness''. The authors argue that the degree to which two industries require similar skills can be inferred from cross-industry labor flows. In essence, two industries are deemed \emph{skill related} if labor flows between them are surprisingly large, compared to a benchmark in which workers move randomly among industries.

In the context of Marshallian agglomeration forces, skill relatedness offers a natural way to determine which industries draw from the same pool of labor, or from the same ``skill basin'' \parencite{oclery_kinsella2022rp}. Moreover, worker mobility is an important vehicle for knowledge transfer, as evidenced by productivity growth and spillovers following labor flows and coworker networks \parencite{lengyel_eriksson2017jeg, eriksson_lengyel2019eg, csafordi_et_al2020tjtt}. Therefore, apart from identifying skill basins, skill relatedness is also likely to be a good proxy for cognitive proximity between industries.

\subsection{Value chains and knowledge transfer} 
\label{sec:subsec_value_chains_knowledge}

The importance of local interactions and the cluster literature's emphasis on localized buyer-supplier networks would, \emph{prima facie}, seem at odds with the rapid growth of GVCs over the past decades \parencite{gereffi2005govgvc, baldwin2022gvc, johnson2018gvc}. That is, the existence of GVCs suggest that improvements in transportation and communication technologies have allowed coordinating buyer-supplier interactions over long distances. However, this does not hold true for all parts of GVCs: while the production and assembly activities at the middle of the value chain have become geographically mobile, high value-added activities at both ends of the so-called smile curve \parencite{baldwin_ito2021cje}, like R\&D and design, or marketing and associated services, exhibit substantial spatial (co-)concentration \parencite{mudambi2008location}.

One of the characteristics that these spatially sticky elements of GVCs share is their reliance on tacit knowledge. Because tacit knowledge is so hard to transmit over long distances \parencite{jaffe1993spillovers, audretsch1996spillovers}, buyer-supplier relations that embed much tacit knowledge will benefit from geographical proximity. This point is well-established in the literature on regional innovation systems \parencite[e.g.,][]{cooke_morgan1994ijtm}. For instance, in some cases, innovation needs to be coordinated along the value chain \parencite{azadegan2010supplierinnov}, impelling suppliers to work with their customers to integrate new technologies or help improve end products. Such interactions require knowledge transfers and learning processes that are facilitated by geographical proximity \parencite{cooke1996sbe}. In other instances, intermediate goods can be used without much knowledge of how they are made, which arguably allows for more spatial separation between value chain partners. While buyer-supplier relationships thus vary widely in the extent to which they need to embed -- often highly tacit -- knowledge, this diversity has not been explicitly considered in the literature on coagglomeration.

\subsection{Main hypothesis}
\label{sec:subsec_hypothesis}

Based on these different bodies of research, we expect that Marshallian coagglomeration forces will reinforce each other. In particular, not all buyer-supplier linkages require spatial proximity. Instead, we hypothesize that buyer-supplier linkages will only require that firms colocate if the transactions also involve transferring knowledge. Furthermore, we expect that industries that share the same pool of labor are often also cognitively proximate. As a consequence, measures that assess the potential for labor pooling, such as labor flow based skill relatedness, often also shed light on the degree to which industries operate in technologically similar environments. 

Taken together, these arguments lead to the main hypothesis in this paper: \emph{input-output connections enhance industries to coagglomerate only if the industries are also skill related}. We test this hypothesis in the context of industrial coagglomeration in Hungary by first adopting the approach proposed by \textcite{egk2010coagglom} in our data, and then unpack the interacting roles of input-output and skill relatedness at the aggregate level of industry pairs, as well as at the micro-level of firm-to-firm transactions.

\section{Empirical setting} \label{sec:setting}

\subsection{Measuring the coagglomeration of industries}\label{sec:coagglomertaion}

Our empirical work relies on a firm-level dataset, made available by the Hungarian Central Statistical Office (HCSO) through the Databank of the HUN-REN Centre for Economic and Regional Studies\footnote{Databank of HUN-REN Centre for Economic and Regional Studies, https://adatbank.krtk.mta.hu/en/} (HUN-REN CERS Databank). It contains information from balance sheets of companies doing business in Hungary. The data include the location of the company seats (headquarter) at the municipal level, the main activity of the firms as four-digit NACE codes (Statistical Classification of Economic Activities in the European Community, NACE Rev. 2 classification), the number of employees and further balance sheet indicators. We focus on 2017 as this is the year for which all of these datasets provide information. A detailed description of the firm level data and the distribution of employment and firms across regions and industries can be found in Section 1 of the Supplementary information.

To measure the degree to which firms from different industries tend to colocate, we focus on companies with at least two employees and aggregate the firm level employment data to an industry-region matrix. We use this matrix to quantify the tendency of industry $i$ to coagglomerate with industry $j$, using the following metric proposed by \textcite{egk2010coagglom}:

\begin{equation}
\label{eq:eq_egk}
    EGK_{ij} = \frac{\sum_{r=1}^{R} ({s}_{ir} - {x}_{r})({s}_{jr} - {x}_{r})}{1 - \sum_{r=1}^{R} x_{r}^2}
\end{equation}

\noindent
where $s_{ir} = E_{ir} / E_{r}$ is the employment share of industry $i$ in region $r$ (missing indices indicate summations such as $E_{r}=\sum_{r=1}^{R}E_{ir}$), while $x_{r}$ is the mean of these shares in region $r$ across all industries. \textcite{eg1997concentration, eg1999concentration} show that this index quantifies the likelihood that firms in industries $i$ and $j$ generate spillovers for each other in a simple location choice model. The index is widely adopted and used as a benchmark (see e.g., \cite{diodato2018coagglom, steijn2022dynamics, juhasz2021relatedness}), as it is largely independent of the distribution of firm sizes in industries and the granularity of spatial units. We calculate this index for both NUTS3 and NUTS4 regions in Hungary for all pairs of three-digit industries and hereafter refer to it as \textit{EGK coagglomeration}. 

As an alternative, inspired by the measures of \textcite{porter2003}, we use the correlation of revealed comparative advantage ($RCA$) vectors to quantify the coagglomeration of industry pairs. This indicator, which we refer to in the following as \textit{LC coagglomeration}, is calculated as follows. First, we calculate the $RCA$ of industries in regions:

\begin{equation}
\label{eq:eq_rca}
    RCA_{ir}=(E_{ir}/E_{r})/(E_{i}/E)
\end{equation}

\noindent
A region is specialized in an industry when its $RCA$ value is above 1. Next, we use the $RCA$ values to create a binary specialization matrix $M_{ir}$:

\begin{equation}
\label{eq:eq_binary}
    M_{ir} = 
\left\{
\begin{array}{ccl}
    1 & \textrm{if} & RCA_{ir} >= 1 \\
    0 & \textrm{if} & RCA_{ir} < 1
\end{array}
\right.
\end{equation}

\noindent
Finally, we calculate the LC coagglomeration of two industries as the correlation between industries' specialization vectors:

\begin{equation}
\label{eq:eq_coagg_porter}
    LC_{ij} = corr({m}_{i}, {m}_{j}),
\end{equation}

\noindent
where $m_i$ and $m_j$ are column vectors of matrix $M$ that describe the spatial distributions of industries $i$ and $j$.

\begin{figure}[!ht]
\includegraphics[width=0.95\textwidth]{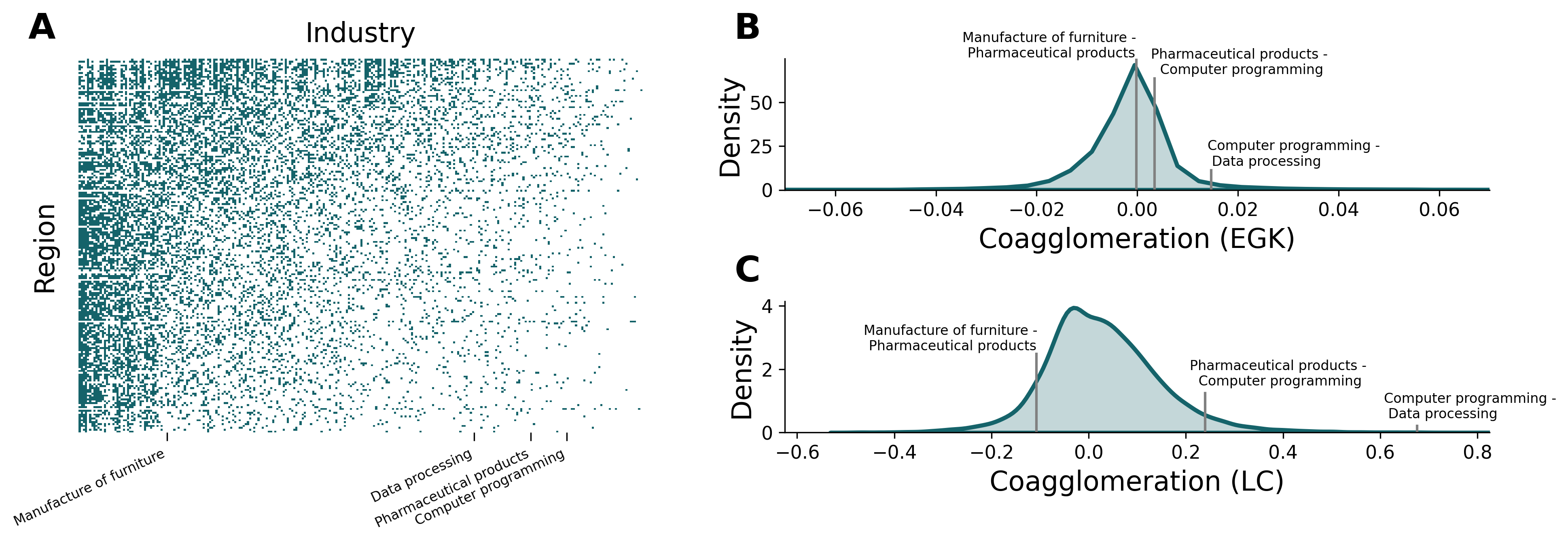}  
\caption{Constructing coagglomeration measures from a region-industry employment matrix.
\textbf{(A)} Region-industry matrix based on NUTS3 regions and three-digit NACE codes.
\textbf{(B)} The distribution of EGK coagglomeration values. For this illustration the tail of the distribution with extreme values was cut off. The unedited figure can be found in Section 2 of the Supplementary information.
\textbf{(C)} The distribution of LC coagglomeration values.}
\label{fig:coagglomeration_illustration}
\end{figure}

We calculate this index on the basis of both NUTS3 and NUTS4 regions in Hungary for all pairs of three-digit industries. The above specifications are two prominent ones among the many ways to calculate coagglomeration indicators \parencite{li2023evaluating}. Figure~\ref{fig:coagglomeration_illustration} illustrates the construction of these measures from a region-industry matrix, yielding plausible variation in the coagglomeration intensity of various industry pairs. In Section 2 of the Supplementary information we provide detailed descriptive statistics on both of these dependent variables and compare them to common alternatives.

\subsection{Skill relatedness}\label{sec:SR}

Previous studies established a number of approaches to capture the extent to which two industries can draw from the same labor pool, including comparing the occupational composition of industries (e.g., \cite{egk2010coagglom, diodato2018coagglom}), and measuring significant labor flows between them (e.g., \cite{neffke2013skillrel, neffke2017interind}). In this study we opt for the latter approach where the central assumption is that workers tend to switch jobs between industries across which they can transfer most of their  skills and human capital. 

To do so we rely on a Hungarian matched employer-employee dataset managed by the HUN-REN CERS Databank. This longitudinal dataset contains the work history of a randomly selected 50\% of the total population on a monthly basis between 2003 and 2017. It links data from different registers, including the Pension Directorate, the Tax Office, the Health Insurance Fund, the Office of Education, and the Public Employment Service, thereby providing comprehensive information on workers and their employers. Unique and anonymized identifiers for both individuals and firms allow us to track the transition of individuals between firms. We use all observed employment spells for each individual in the dataset to establish employee transitions from one firm to the next. In cases when an individual had multiple parallel employment spells before switching, we consider labor flow ties to be created between the new employer and each of the previous employers. This information on monthly labor flows between firms is then pooled across 2015-2017.

To measure the skill relatedness of industry pairs, we aggregate firm-to-firm labor flows to the industry-industry level. Following the approach of \textcite{neffke2013skillrel} and \textcite{neffke2017interind}, the skill relatedness between two three-digit industries (NACE Rev. 2 classification) ($i$ and $j$) is measured by comparing the observed labor flow between them ($F_{ij}$) with what would be expected based on their propensity to take part in labor flows ($(F_{i}F_{j})/F_{}$).

\begin{equation}
\label{eq:eq_labor_flow}
    SR_{ij} = \frac{F_{ij}}{(F_{i}F_{j})/F}
\end{equation}

\noindent
Here, $F_{i}$ is the total outflow of workers from industry $i$, $F_{j}$ is the total inflow to $j$ and $F$ is the total flow of workers in the system. Next, we take the average of $SR_{ij}$ and $SR_{ji}$ to obtain a symmetric measure. Finally, due to the asymmetric range of the measure ($[0,\infty)$), we normalize it between $-1$ and $+1$ ($\tilde{SR}_{ij} = \frac{SR_{ij} - 1}{SR_{ij} + 1}$) (see \cite{neffke2017interind}). As a result, positive values of the final skill relatedness measure correspond to larger-than-expected labor flows. Further details can be found in Section 2 of the Supplementary information.

\subsection{Input-output relations}\label{sec:IO}

To assess the input-output similarity of industries, we rely on two different types of datasets. First, we rely on data from the World Input-Output Database (WIOD). Using the input-output table for Hungary \parencite{timmer2015wiod} for 2014, we obtain directed buyer-supplier relations between two-digit industries.

Using WIOD tables makes our analysis comparable to previous studies, which have also relied on aggregate input-output tables \parencite{egk2010coagglom, diodato2018coagglom}. However, these country-level aggregates may hide much important detail. Therefore, we use a second, micro-level dataset that records business transactions between companies in Hungary. These data are derived from the value added tax (VAT) reports collected by the National Tax and Customs Administration of Hungary. Firms are obligated to declare all business transactions in Hungary if the VAT content of their operations exceeds ca. 10,000 EUR in that year. The dataset is anonymized and available for research purposes through the HUN-REN CERS Databank. It has been used to construct interfirm supplier networks to study production processes, systemic risks and interdependencies between companies at the national scale \parencite{diem2022quantifying, lorincz2023transactions, pichler2023science}.

We aggregate firm-to-firm supplier transaction values between 2015 and 2017 to the level of pairs of three-digit industries to derive a dataset that is similar in structure to the IO tables in the WIOD data. However, we will also use the micro data themselves to analyze colocation at the firm level.

To construct an indicator that captures the strength of value chain linkages between two industries, we follow the same approach as for skill relatedness in eq. (\ref{eq:eq_labor_flow}): 

\begin{equation}
\label{eq:eq_io}
    IO_{ij} = \frac{V_{ij}}{(V_{i}V_{j})/V}
\end{equation}

\noindent
where $V_{ij}$ stands for the total value of goods and services that industry $i$ supplies to industry $j$. Furthermore, omitted indices indicate summations over the corresponding dimensions. As before,  the ratio compares observed flows to expected flows. We once again symmetrize the index, taking the average of $IO_{ij}$ and $IO_{ji}$, and then use the same rescaling as for skill relatedness to map all values between -1 and +1 resulting in $\tilde{IO}_{ij}$. 

We calculate this index at the two-digit level using WIOD data (\textit{IO (WIOD)}) and at the three-digit level using the aggregated transaction values from the VAT records (\textit{IO (transactions)}). Basic descriptive statistics for both measures are provided in Table~\ref{tab:descriptive_stats}. More detailed statistics can be found in Section 2 of the Supplementary information.

\begin{table}[!htbp]
\begin{minipage}{\linewidth}
\centering
\fontsize{10}{12}\selectfont
\captionsetup{singlelinecheck=off, skip=0.ex}
\caption{Descriptive statistics} 
\label{tab:descriptive_stats} 
\begin{tabular}{@{\extracolsep{5pt}}lcccc} 
\hline \\[-1.8ex] 
Variable & \multicolumn{1}{c}{Mean} & \multicolumn{1}{c}{Std. dev.} & \multicolumn{1}{c}{Min} & \multicolumn{1}{c}{Max}  \\ 
\hline \\[-1.8ex] 
Coagglomeration (EGK) NUTS3 & $-$0.001 & 0.077 & $-$0.319 & 0.966\\ 
Coagglomeration (LC) NUTS3 & 0.031 & 0.282 & $-$1.000 & 1.000 \\ 
Coagglomeration (EGK) NUTS4 & $-$0.001 & 0.017 & $-$0.071 & 0.762\\ 
Coagglomeration (LC) NUTS4 & 0.029 & 0.119 & $-$0.488 & 0.807\\ 
Labor (SR) & $-$0.432 & 0.495 & $-$1.000 & 0.999 \\ 
IO (WIOD) & $-$0.520 & 0.449 & $-$1.000 & 0.857 \\ 
IO (transactions) & $-$0.691 & 0.447 & $-$1.000 & 0.999 \\ 
\hline \\[-1.8ex] 
\end{tabular} 
\caption*{\textbf{Note:} Statistics for all variables are calculated from 35778 observations.}
\end{minipage}
\end{table}

\section{Results} \label{sec:result}

\subsection{Drivers of coagglomeration in Hungary}

We start our analysis with a replication of the findings of \textcite{egk2010coagglom} for Hungary. Following \textcite{diodato2018coagglom}, we focus on the labor pooling and input-output channels. To assess the relative importance of either channel as a driver of coagglomeration patterns, we estimate the following baseline equation, using Ordinary Least Squares (OLS) regression:

\begin{equation}
\label{eq:eq_ols}
    Coagglomeration_{ij} = \beta_0 + \beta_1\tilde{SR}_{ij} + \beta_2\tilde{IO}_{ij} + \epsilon_{ij}
\end{equation}

\noindent
where $\tilde{SR}_{ij}$ and $\tilde{IO}_{ij}$ refer to the skill relatedness and input-output dependency measures defined in sections \ref{sec:SR} and \ref{sec:IO}. The dependent variable, $Coagglomeration_{ij}$, is either the EGK or the LC coagglomeration index described in section \ref{sec:coagglomertaion}. 

Unlike \textcite{egk2010coagglom}, who only consider manufacturing industries, and \textcite{diodato2018coagglom}, who compare manufacturing and service industries, our analysis includes all sectors of the economy. Moreover, we run our analysis twice, where the coagglomeration of industries is either measured within NUTS3 regions or within NUTS4 regions. Our preferred specifications are those based on NUTS4 regions, where Budapest, the capital of Hungary, is divided into 23 micro-regions. 

Table~\ref{tab:ols_coagglomeration} presents the OLS regression results. To facilitate the interpretation of the effect sizes, we rescale all variables such that they are expressed in units of standard deviations. This rescaling is applied to all subsequent analyses. Results are qualitatively in line with those in \textcite{egk2010coagglom} and \textcite{diodato2018coagglom}: also in Hungary, both Marshallian channels are significant drivers of coagglomeration. Moreover, labor pooling seems to play a more important role than input-output connections. The exception to this is when we measure coagglomeration using locational correlations and input-output linkages are based on actual firm-to-firm transactions. In this case, labor pooling and value chain connections contribute about equally to the coaglomeration patterns we observe.

\begin{table}[!htbp]
\begin{minipage}{\linewidth}
\centering
\fontsize{10}{12}\selectfont
\captionsetup{singlelinecheck=off, skip=0.ex}
\caption{OLS multivariate regressions} 
\label{tab:ols_coagglomeration} 
\begin{tabular}{@{\extracolsep{1pt}}lcccccccc} 
\\[-1.8ex]\hline 
\\[-1.8ex] & \multicolumn{4}{c}{Coagglomeration (EGK)} & \multicolumn{4}{c}{Coagglomeration (LC)} \\  
\\[-2.5ex] & NUTS3 & NUTS4 & NUTS3 & NUTS4 & NUTS3 & NUTS4 & NUTS3 & NUTS4\\ 
\\[-2.5ex] & (1) & (2) & (3) & (4) & (5) & (6) & (7) & (8)\\ 
\hline \\[-1.8ex]
Labor (SR) & 0.096$^{***}$ & 0.061$^{***}$ & 0.084$^{***}$ & 0.057$^{***}$ & 0.142$^{***}$ & 0.175$^{***}$ & 0.106$^{***}$ & 0.126$^{***}$\\
& (0.019) & (0.012) & (0.020) & (0.012) & (0.018) & (0.024) & (0.016) & (0.020)\\   
IO (WIOD) & 0.056$^{***}$ & 0.041$^{***}$ & & & 0.064$^{***}$ & 0.089$^{***}$ & & \\   
& (0.018) & (0.011) & & & (0.021) & (0.022) & & \\   
IO (transactions) & & & 0.050$^{**}$ & 0.026$^{*}$ & & & 0.112$^{***}$ & 0.153$^{***}$\\   
 & & & (0.025) & (0.015) & & & (0.018) & (0.020)\\ 
\hline \\[-1.8ex] 
Observations & 35,778 & 35,778 & 35,778 & 35,778 & 35,778 & 35,778 & 35,778 & 35,778 \\ 
R$^2$ & 0.014 & 0.006 & 0.013 & 0.005 & 0.027 & 0.043 & 0.033 & 0.055\\ 
Adjusted R$^2$ & 0.014 & 0.006 & 0.013 & 0.005 & 0.027 & 0.043 & 0.033 & 0.055\\ 
\hline \\[-1.8ex]
\end{tabular} 
\caption*{\textbf{Note:} Clustered (industry$_i$ and industry$_j$) standard errors in parentheses.  Significance codes: ***: p$<$0.01, **: p$<$0.05, *: p$<$0.1.}
\end{minipage}
\end{table}

\textcite{egk2010coagglom} raise the concern that  coagglomeration patterns may not only be a consequence of labor pooling and value chain linkages, but also cause these linkages themselves. For instance, industries may use similar labor because they are colocated, not vice versa. Similarly, industries may preferentially use inputs that are available nearby and adjust their production technologies accordingly instead of value chain links causing firms to coagglomerate.\footnote{As an example, \textcite{egk2010coagglom} point to the trade between shoe manufacturers and leather producers. The volume of this trade  may reflect more than just the inherent technological features of shoe manufacturing: shoes can be made out of several materials, including leather, but also plastics. The choice of leather as an input to shoe-making may therefore be a consequence of an idiosyncratic historical colocation of leather producers with shoe producers. Similarly, shoe manufacturers may have hired workers, not only by their  suitability for shoe-making, but also according to their availability on the the local labor market. In the longer run, shoe makers may have adjusted their production processes to make better use of these locally available workers. This would once again lead to some reverse causality between the linkages between industries and their coagglomeration patterns.} To address this, the authors instrument the different types of linkages between industries with analogous quantities calculated from data for economies other than the US.

\begin{table}[!htbp]
\begin{minipage}{\linewidth}
\centering
\fontsize{10}{12}\selectfont
\captionsetup{singlelinecheck=off, skip=0.ex}
\caption{Labor channel through instrumental variable univariate regressions} 
\label{tab:iv_univar_labor}
\begin{tabular}{@{\extracolsep{1pt}}lcccc} 
\\[-1.8ex]\hline 
\\[-1.8ex] & \multicolumn{2}{c}{Coagglomeration (EGK)} & \multicolumn{2}{c}{Coagglomeration (LC)} \\  
\\[-2.5ex] & NUTS3 & NUTS4 & NUTS3 & NUTS4\\ 
\\[-2.5ex] & (1) & (2) & (3) & (4)\\ 
\hline \\[-1.8ex]
Labor (SR) & 0.403$^{***}$ & 0.285$^{***}$ & 0.417$^{***}$ & 0.575$^{***}$ \\
& (0.074) & (0.045) & (0.066) & (0.083) \\ 
\hline \\[-1.8ex] 
Observations & 35,778 & 35,778 & 35,778 & 35,778 \\ 
R$^{2}$ & -0.078 & -0.043 & -0.048 & -0.114 \\ 
Adjusted R$^{2}$ & -0.078 & -0.043 & -0.048 & -0.114 \\
KP F-statistic & 115.955 & 115.955 & 115.955 & 115.955 \\ 
\hline \\[-1.8ex]
\end{tabular}
\caption*{\textbf{Note:} Clustered (industry$_i$ and industry$_j$) standard errors in parentheses.  Significance codes: ***: p$<$0.01, **: p$<$0.05, *: p$<$0.1.}
\end{minipage}
\end{table}

In Tables~\ref{tab:iv_univar_labor} and \ref{tab:iv_univar_io}, we follow the same identification strategy. To instrument our labor market pooling variable, we construct a skill relatedness measure between three-digit industries using data from the Swedish labor market. To instrument value chain linkages we average all input-output tables in the WIOD data, excluding the Hungarian tables. Detailed description on the instruments can be found in Section 2 of the Supplementary information. Our instruments are valid, as long as idiosyncratic patterns in the input-output and labor dependencies outside Hungary are exogenous to coagglomeration patterns inside Hungary. As in \textcite{egk2010coagglom}, we run univariate analyses, testing for the causal effect of each channel separately. Univariate OLS regressions for comparison are provided in Section 3 of the Supplementary information, while in Section 4 we present the first and second stages of instrumental variable estimations separately.

\begin{table}[!htbp]
\begin{minipage}{\linewidth}
\centering
\fontsize{10}{12}\selectfont
\captionsetup{singlelinecheck=off, skip=0.ex}
\caption{Input-output channel through instrumental variable univariate regressions} 
\label{tab:iv_univar_io}
\begin{tabular}{@{\extracolsep{1pt}}lcccccccc} 
\\[-1.8ex]\hline 
\\[-1.8ex] & \multicolumn{4}{c}{Coagglomeration (EGK)} & \multicolumn{4}{c}{Coagglomeration (LC)} \\  
\\[-2.5ex] & NUTS3 & NUTS4 & NUTS3 & NUTS4 & NUTS3 & NUTS4 & NUTS3 & NUTS4\\ 
\\[-2.5ex] & (1) & (2) & (3) & (4) & (5) & (6) & (7) & (8)\\ 
\hline \\[-1.8ex]
IO (WIOD) & 0.083$^{***}$ & 0.054$^{***}$ & & & 0.083$^{***}$ & 0.106$^{***}$ & & \\
& (0.020) & (0.012) & & & (0.023) & (0.023) & & \\ 
IO (transactions) & & & 0.475$^{***}$ & 0.310$^{***}$ & & & 0.479$^{***}$ & 0.609$^{***}$ \\
 & & & (0.129) & (0.078) & & & (0.128) & (0.141)\\
\hline \\[-1.8ex] 
Observations & 35,778 & 35,778 & 35,778 & 35,778 & 35,778 & 35,778 & 35,778 & 35,778 \\ 
R$^{2}$ & 0.005 & 0.002 & -0.146 & -0.066 & 0.007 & 0.013 & -0.081 & -0.122\\
Adjusted R$^{2}$ & 0.005 & 0.002 & -0.146 & -0.066 & 0.007 & 0.013 & -0.081 & -0.122\\
KP F-statistic & 12000 & 12000 & 38.276 & 38.276 & 12000 & 12000 & 38.276 & 38.276 \\
\hline \\[-1.8ex]
\end{tabular}
\caption*{\textbf{Note:} Clustered (industry$_i$ and industry$_j$) standard errors in parentheses.  Significance codes: ***: p$<$0.01, **: p$<$0.05, *: p$<$0.1.}
\end{minipage}
\end{table}

Our results corroborate the OLS analysis of Table~\ref{tab:ols_coagglomeration}: both channels have a large and causal effect on coagglomeration patterns in Hungary. Moreover, the labor pooling channel has a stronger causal effect than input-output relations, unless we measure the input-output relations using micro-level transaction data. In the latter case, labor and value chains represent about equally strong coagglomeration forces. 

These results strengthen the external validity of the literature on coagglomeration, which has mostly focused on the US economy. The generalizability of those results to Hungary is not trivial: the Hungarian economy is much smaller than the US economy. This not only affects the amount of spatial variation that is available for our estimations, but also the degree to which firms can rely on domestic value chains. In the Supplementary information, we show multiple robustness checks of these results. 

In Section 5 of the Supplementary information we present results of using alternative instruments for input-output connections such as US supply tables or the WIOD table for the Czech Republic only. Section 6 of the Supplementary information presents the above OLS and IV regressions separately for manufacturing and service industries. In Section 7, we apply geographical restrictions and exclude firms located in Budapest from our sample. Furthermore, we re-run our main models focusing only on firms with a single plant location. Our main results are valid for all but a few of the listed specifications.

Finally, we also try estimating multivariate IV regressions, where both agglomeration forces enter the regression simultaneously. However, the Kleibergen-Paap statistic for these analyses indicates that these models typically suffer from weak instruments. That is, there is insufficient variation available in our data to reliably disentangle the causal effects of labor pooling and value chain linkages. Results of these models are reported in Section 8 of the Supplementary information.

\subsection{Interaction effects}\label{sec:interaction}

We now turn to testing our main hypothesis that the role of value chain links in determining the locations of industries depends on the degree to which industries are also skill related. To do so, we interact the metrics for the two Marshallian channels in the following model:

\begin{equation}
\label{eq:eq_interact}
    {Coagglomeration}_{ij} = \tilde{\beta}_0 + \tilde{\beta}_1\tilde{SR}_{ij} + \tilde{\beta}_2\tilde{IO}_{ij} + \tilde\beta_{12}\tilde{IO}_{ij}\tilde{SR}_{ij} + \eta_i + \delta_j +\tilde{\epsilon}_{ij}
\end{equation}

Because our instruments proved too weak to estimate multivariate models, these models are estimated using OLS regressions. To nevertheless minimize confounding, we add two-way industry fixed effects, denoted by $\eta_i$ and $\delta_j$.

\begin{table}[!htbp]
\begin{minipage}{\linewidth}
\centering
\fontsize{10}{12}\selectfont
\captionsetup{singlelinecheck=off, skip=0.ex}
\caption{OLS multivariate regressions with interaction effects} 
\label{tab:interactions} 
\begin{tabular}{@{\extracolsep{0.75pt}}lcccccccc} 
\\[-1.8ex]\hline 
\\[-1.8ex] & \multicolumn{4}{c}{Coagglomeration (EGK)} & \multicolumn{4}{c}{Coagglomeration (LC)} \\ 
\\[-2.5ex] & NUTS4 & NUTS4 & NUTS4 & NUTS4 & NUTS4 & NUTS4 & NUTS4 & NUTS4\\ 
\\[-2.5ex] & (1) & (2) & (3) & (4) & (5) & (6) & (7) & (8) \\ 
\hline \\[-1.8ex] 
Labor (SR) & 0.059$^{***}$ & 0.052$^{***}$ & 0.057$^{***}$ & 0.051$^{***}$ & 0.173$^{***}$ & 0.164$^{***}$ & 0.126$^{***}$ & 0.138$^{***}$\\ 
& (0.012) & (0.013) & (0.011) & (0.013) & (0.024) & (0.016) & (0.020) & (0.013)\\  
IO (WIOD) & 0.037$^{***}$ & 0.059$^{***}$ & & & 0.084$^{***}$ & 0.057$^{***}$ & & \\
& (0.011) & (0.018) & & & (0.022) & (0.020) & & \\
IO (WIOD)*Labor & 0.023$^{***}$ & 0.016$^{**}$ & & & 0.034$^{**}$ & 0.033$^{***}$ & & \\ 
& (0.008) & (0.008) & & & (0.015) & (0.010) & & \\
IO (trans) & & & 0.012 & 0.018$^{*}$ & & & 0.119$^{***}$ & 0.085$^{***}$\\ 
& & & (0.017) & (0.011) & & & (0.021) & (0.013) \\ 
IO (trans)*Labor & & & 0.025$^{***}$ & 0.019$^{**}$ & & & 0.065$^{***}$ & 0.051$^{***}$\\   
& & & (0.010) & (0.009) & & & (0.013) & (0.008)\\
\hline \\[-1.8ex] 
\\[-2.5ex]Two way FE & No & Yes & No & Yes & No & Yes & No & Yes \\ 
\\[-2.5ex]Observations & 35,778 & 35,778 & 35,778 & 35,778 & 35,778 & 35,778 & 35,778 & 35,778 \\ 
R$^{2}$ & 0.007 & 0.072 & 0.006 & 0.072 & 0.045 & 0.317 & 0.060 & 0.325\\  
Adjusted R$^{2}$ & 0.007 & 0.058 & 0.006 & 0.058 & 0.045 & 0.307 & 0.060 & 0.315\\ 
\hline \\[-1.8ex] 
\end{tabular} 
\caption*{\textbf{Note:} Clustered (industry$_i$ and industry$_j$) standard errors in parentheses.  Significance codes: ***: p$<$0.01, **: p$<$0.05, *: p$<$0.1.}
\end{minipage}
\end{table}

Table~\ref{tab:interactions} shows results for our preferred specification based on coagglomeration in NUTS4 regions. The interaction effects are positive in all specifications, corroborating our hypothesis that labor pooling and value chain effects reinforce each other.

\begin{figure}[!ht]
\includegraphics[width=0.95\textwidth]{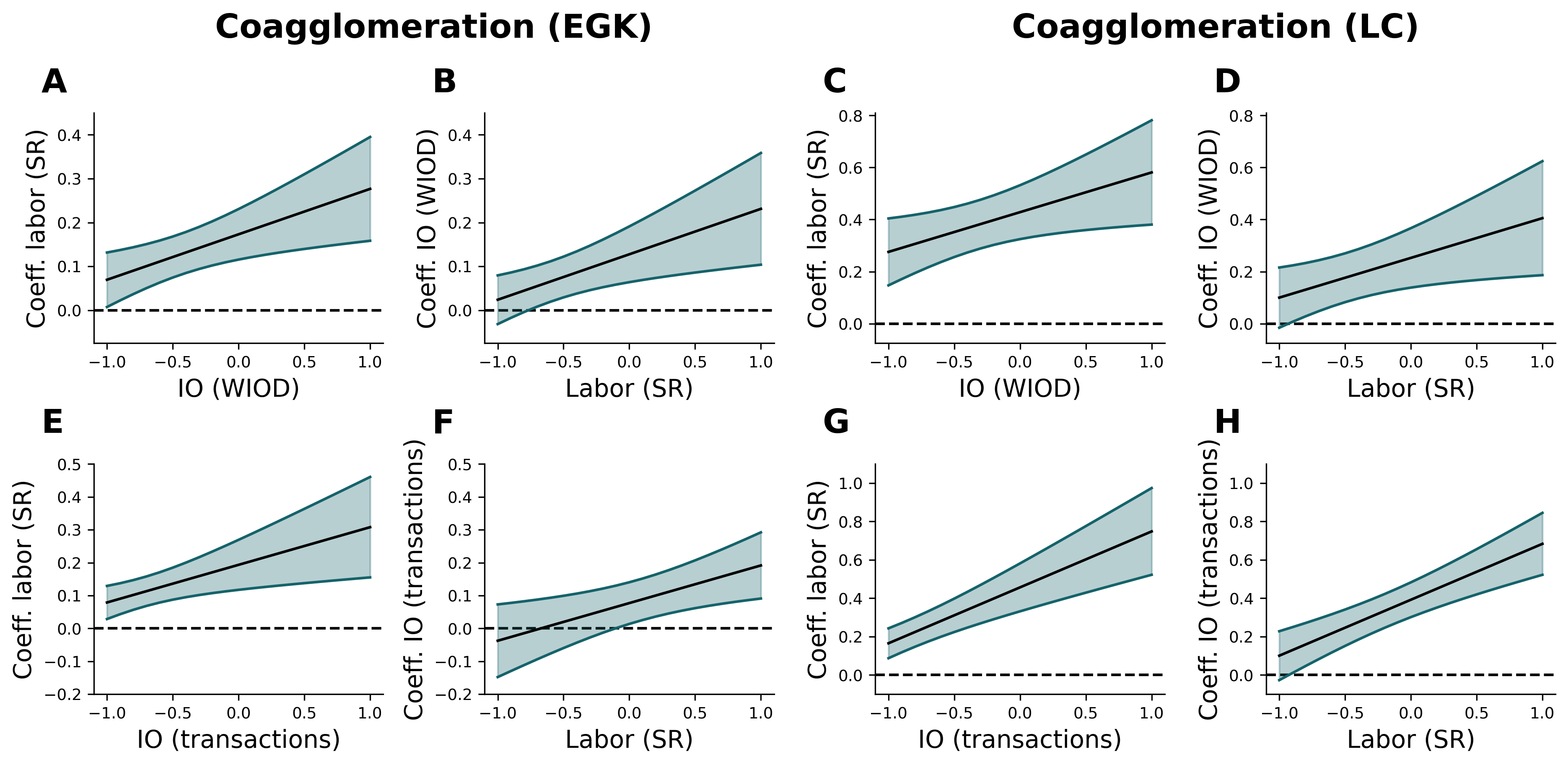}  
\caption{Reinforcing effect of input-output connections measured by IO (WIOD) and labor flow on coagglomeration.
\textbf{(A)} The influence of labor flow on Coagglomeration (EGK) at different levels of IO (WIOD) connections.
\textbf{(B)} The influence of IO (WIOD) connections on Coagglomeration (EGK) at different levels of labor flow. 
\textbf{(C)} The influence of labor flow on Coagglomeration (LC) at different levels of IO (WIOD) connections.
\textbf{(D)} The influence of IO (WIOD) connections on Coagglomeration (LC) at different levels of labor flow.
\textbf{(E)}, \textbf{(F)}, \textbf{(G)}, \textbf{(H)} are based on the same settings as the upper row, but IO connections are measured through the transaction data.
Visualizations are based on Table~\ref{tab:interactions} Model (1), (3), (5) and (7). Green areas depict 95\% confidence intervals.}
\label{fig:interplots_main}
\end{figure}

Figure~\ref{fig:interplots_main} visualizes the implied effects of labor pooling for different values of input-output linkages in panels A, C, E and G. Along the vertical axis, these graphs plot the effect of labor pooling on coagglomeration,  $\tilde{\beta}_1 + \tilde{\beta}_{12} \tilde{IO}_{ij}$, for varying levels of value chain connections between industries $i$ and $j$,  $\tilde{IO}_{ij}$. Note that the range of the horizontal axis is limited to the values that $\tilde{IO}_{ij}$ can theoretically attain. These panels show that labor pooling effects are positive and significant at any level of value chain linkages.

This contrasts with the effect of value chain linkages, $\tilde{\beta}_2 + \tilde{\beta}_{12} \tilde{SR}_{ij}$, shown in panels B, D, F and H. The value chain effect is in general positive, but drops to zero when skill relatedness between industries equals -1, which happens in 35\% of all industry combinations. In other words, value chain partners only tend to significantly coagglomerate if they are not completely unrelated in terms of the skills of their workforces. 

These results hold regardless of whether we use the EGK or LC measures of coagglomeration and whether we measure value chain linkages using WIOD data or derive them from micro-level transaction data. In Section 9 of the Supplementary information, we show results for various other regression specifications with the interaction term.

\subsection{Firm-to-firm ties behind coagglomeration}

Our data allow us to analyze not just the potential connections between industries that have been commonly studied in the coagglomeration literature, but the actual connections between firms. We can do so both in terms of transactions and of labor flows. This allows us to create networks of firms that are connected either if they supply (or purchase) goods or services, or if workers move from one firm to the other. To simplify the analysis, we consider ties as undirected and unweighted in both networks.   

Table~\ref{tab:netdescr} provides a general description of these networks for the period 2015-2017. Overall, the input-output network has a lower number of connections, which may in part reflect VAT reporting thresholds \parencite{pichler2023science}. In addition, it is less transitive than the labor flow network, which is consistent with previous findings that show that production networks have fewer closed triangles than other social networks \parencite{mattsson2021functional}. When it comes to geography, the descriptive statistics of Table~\ref{tab:netdescr} suggests that labor flows are more spatially concentrated than transaction flows. While 48\% of the observed labor flows between firms take place within the same NUTS3 region, only 41\% of the transaction linkages are intra-regional. Finally, overlapping connections, i.e., pairs of firms that exchange both workers and goods or services are rare but highly concentrated geographically.

\begin{table}[!htbp]
\begin{minipage}{\linewidth}
\centering
\fontsize{10}{12}\selectfont
\captionsetup{singlelinecheck=off, skip=0.ex}
\caption{Descriptive statistics of the labor flow and supplier networks} 
\label{tab:netdescr} 
\begin{tabular}{@{\extracolsep{1pt}}lccc}
\\[-1.8ex]\hline 
\\[-1.8ex] & IO & Labor & IO and Labor \\
\\[-2.5ex]\hline 
\\[-1.8ex]Firms connected & 72445 & 115519 & 16719 \\
\\[-2.5ex]Edges & 194231 & 492818 & 14209 \\
\\[-2.5ex]Average degree & 5.362 & 8.534 & 1.700 \\
\\[-2.5ex]Transitivity & 0.013 & 0.027 & 0.048 \\
\\[-2.5ex]Average distance of ties (km) & 68 & 58 & 40 \\
\\[-2.5ex]Share of edges inside NUTS3 & 41\% & 48\% & 63\% \\
\\[-2.5ex]Share of edges inside NUTS4 & 14\% & 20\% & 41\% \\
\\[-2.5ex]Share of edges inside Budapest & 22\% & 21\% & 25\% \\
\hline \\[-1.8ex] 
\end{tabular}
\caption*{\textbf{Note:} Edges are undirected and unweighted ties that represent any supply or labor exchange between two companies.}
\end{minipage}
\end{table}

\begin{figure}[!ht]
\includegraphics[width=0.95\textwidth]{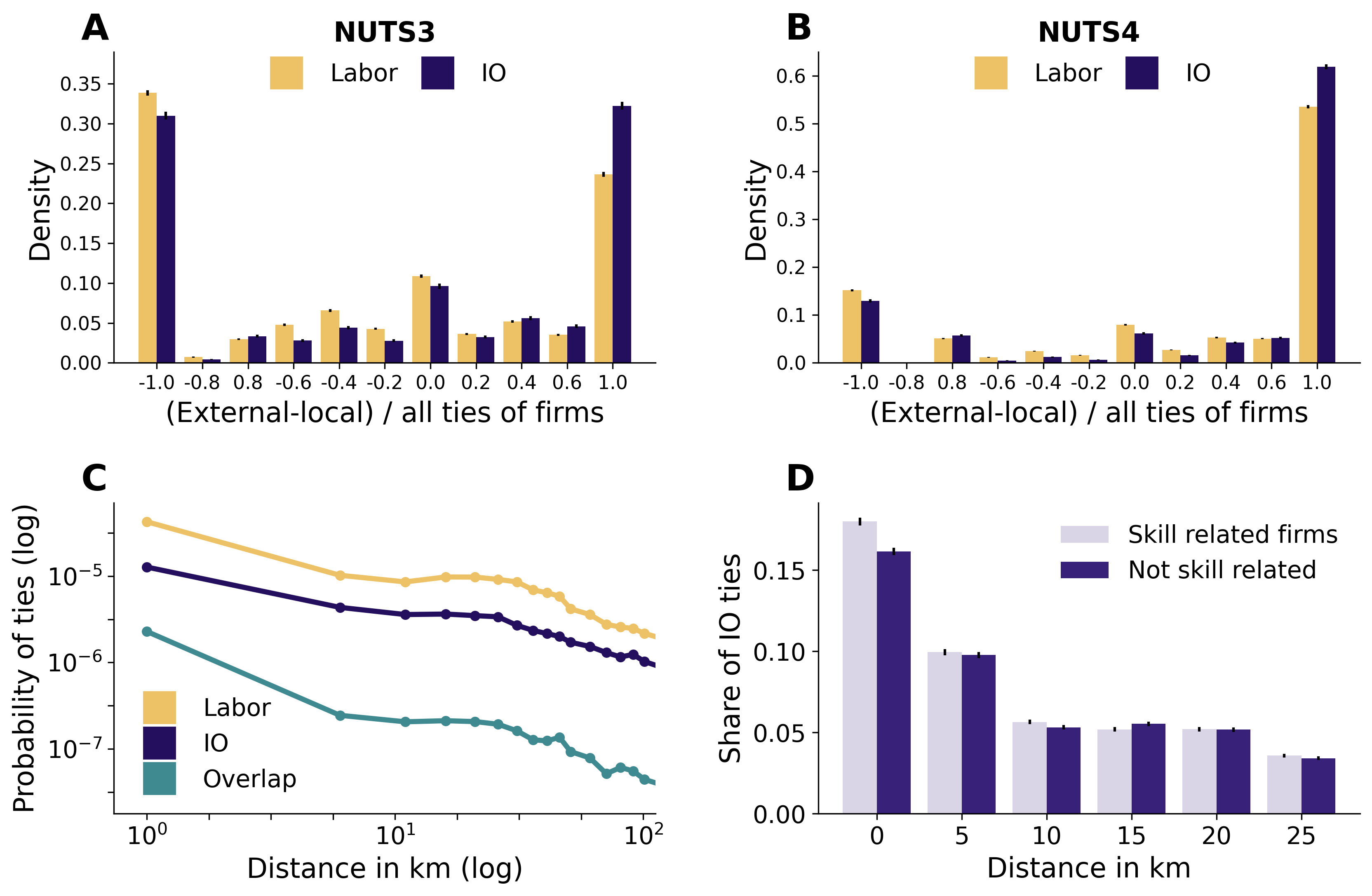} 
\caption{Geography of firm-to-firm input-output (IO) and labor flow connections.
\textbf{(A)} Share of input-output and labor flow ties of firms inside their local NUTS3 region and
\textbf{(B)} inside their local NUTS4 region. 
\textbf{(C)} Probability of labor flow, input-output connections and overlapping (labor and input-output) ties by distance.
\textbf{(D)} Share of input-output connections between firms in skill related (SR) and not skill related industries inside 25 kilometers. The figures are based on the sample of firms we used to construct our aggregate measures.}
\label{fig:distdistr}
\end{figure}

In line with this, Figure~\ref{fig:distdistr}A and B indicate the extent to which actual labor flows and input-output connections happen within or across NUTS3 or NUTS4 regions, respectively. It compares for each firm the number of cross-region (``external'') ties to the number of within-region (``local'') ties the firm maintains, by expressing the difference between them as a share of all ties the firm engages in. The majority of firms either exhibit exclusively local or exclusively interregional ties. This is most visible for NUTS 3 regions, where there is a balance between firms with only local and firms with only external ties. However, what stands out at both spatial scales is that firms tend to have a more local orientation when it comes to their labor flows compared to their transactions with other firms. At more disaggregated level of NUTS4 regions, firms with completely external ties naturally outweigh all other type of firms, as these regions are often part of larger labor markets and agglomeration areas \parencite{toth2014ts}\footnote{Maps of both spatial divisions are provided in Section 1 of the Supplementary information)}.

Figure~\ref{fig:distdistr} further analyzes how interfirm ties decay with the distance between firms. Figure~\ref{fig:distdistr}C plots the likelihood that two firms in Hungary are connected through labor flows, transactions or both for different distance bins. In line with \textcite{bernard2019jpe}, input-output connections are highly concentrated in space. However, labor flow connections are even more sensitive to distance. Within a distance of 10 kilometers, the probability of a labor flow between two firms decreases faster than that of input-output relationships. 

The third line shows the likelihood that two firms are connected in both the labor and the IO network. Note that the figure is plotted using logarithmic axes. Consequently, if the probabilities of being connected in either network were independent, the resulting line should equal the sum of the labor and the IO plots.\footnote{In a log-transformation, multiplications become additions: $\log p_{io}p_{sr} = \log p_{io} + \log p_{sr}$, where $p_{io}$ and $p_{sr}$ are the probabilities that two firms are connected through labor flows or transactions, respectively.} The fact that the Overlap line lies far above this sum means that firms that maintain one connection are disproportionally likely to maintain the other connection as well. Moreover, in line with our finding that labor pooling and value chain linkages reinforce each other's impact on coagglomeration, the Overlap line shows by far the steepest distance decay of all lines.

Finally, Figure~\ref{fig:distdistr}D shows how the share of a firm's transaction ties occur within a given distance. Highlighting this share for distances for up to 25 km, it shows that firm-to-firm transaction ties are more likely to be highly localized if firms belong to skill related industries than if they don't. Apparently, suppliers are more likely to colocate with their buyers (or vice versa) if they operate in industries with high cognitive proximity. Section 10 of the Supplementary information provides further visualizations on firm-to-firm connection patterns.

\section{Conclusion} \label{sec:conclusion}

Why, in a world of globalized supply chains, do supplier-buyer transactions still often take place in close geographical proximity? We propose that this may happen when value chain partners need to exchange not just goods, but also know-how. In that case, spatial proximity facilitates the transfer and coordination of know-how along these value chains. We assess the validity of this proposition using the empirical framework of coagglomeration. Industries coagglomerate whenever their interactions are facilitated by spatial proximity. Consequently, the mere fact that industries belong to the same value chain is insufficient reason for them to coagglomerate. This changes when interactions between value chain partners embed large quantities of tacit knowledge. In that case, value chain relations should reinforce the benefits of colocation.

Knowledge exchange between industries is more likely to occur between industries that are cognitively close to one another. Because of the pivotal role of human capital in firms' competitive advantage, we expect that this cognitive proximity is particularly high between industries that exchange a lot of labor, i.e., that are skill related. We therefore expect that value chain linkages lead to coagglomeration only if industries are also skill related.

We tested this hypothesis by studying the drivers of industrial coagglomeration in Hungary, where detailed datasets from public registers allowed us to complement the traditional coagglomeration framework with a detailed analysis of firm-to-firm networks that underpin coagglomeration. In line with our hypothesis, we find a positive interaction effect between the Marshallian agglomeration channels of labor pooling and input-output linkages. The impact of input-output linkages on coagglomeration increases with the skill relatedness between industries, that is, with the potential to redeploy human capital between them. In fact, if industries only share a value chain connection but no skill relatedness link, we do not find any statistically significant evidence that these industries tend to colocate. 

This result contributes to the growing literature on coagglomeration in several ways. First, whereas the coagglomeration literature has considered various sources of hetereogeneity, it has so far not considered that different Marshallian agglomeration channels may reinforce one another. Second, by replicating the empirical findings of \textcite{egk2010coagglom}, we show that many of the main findings in this and subsequent papers extend beyond the US context. Specifically, using instrumental variables estimation, we replicate the causal effects of input-output linkages and labor pooling on industrial coagglomeration in the small and open economy of Hungary where, unlike in the US, many inputs need to be imported \parencite{halpern_et_al2015aer}, which hinders the formation of long domestic supply chains. Moreover, in line with \textcite{diodato2018coagglom} and \textcite{steijn2022dynamics}, we find that the impact of the labor pooling excedes the impact of value chain linkages.

Second, analyzing actual labor flows and transactions between firms, we show that distance decays are particularly steep in supply-chain connections between firms in cognitively proximate industries. This micro-level evidence lends further plausibility to existing findings about aggregate coagglomeration patterns. Moreover, because coagglomeration patterns are pivotal inputs in the construction of product and industry spaces, these findings may also bear relevance on the literature on economic complexity. This literature argues that economic development unfolds by the gradual accumulation of complementary capabilities in places \parencite{hidalgo2021review, balland2022paradigm}. As capabilities are notoriously difficult to observe directly, the combinatorial potential of places (economic complexity) and capability requirements of activities (activity complexity) are often inferred from the geographical distribution of these activities. In this context, our findings suggest that important localized capabilities reside in value chain interactions that embed tacit knowledge and that such connections are important drivers of the place-activity matrix that underlies industry spaces and complexity metrics. 

Finally, our findings connect to the literature on regional clusters. They highlight  that although value chain connections in a cluster may contribute to the competitiveness of clusters, whether they do so will crucially depend on the extent to which these value chain interactions are enriched with knowledge transfers, possibly facilitated by the exchange of skilled labor.

Our study also has several limitations. First, while Hungary represents a novel test case with detailed information on the drivers of industrial coagglomeration, the country is also strongly dependent on exports and imports. Such dependencies are presently not covered in our data and adding export and import data would be a valuable extension of the current work. In addition, foreign direct investment (FDI) and multinational enterprises (MNEs) play an important role in the Hungarian economy. Moreover, this foreign-owned part of the economy may behave very differently from the domestic economy \parencite{bekes_et_al2009we, halpern_et_al2015aer, elekes_et_al2019rs}, because these firms can access resources in other locations through their internal corporate networks. Distinguishing between coagglomeration patterns of foreign-owned and domestic firms therefore represents a promising avenue for future research. This would connect the research on coagglomeration forces to a well-established literature on knowledge spillovers from MNEs to their host regions, which are often mediated through value chains linkages. 

Second, there are only a few years in which all official registers are available. This precluded analyzing changes in coagglomeration forces over time. In particular, we could not assess whether the stronger colocation of value chain interactions among cognitively proximate partners is a new phenomenon, or has persisted over time. Because the coagglomeration literature has pointed to an increased importance of labor pooling and knowledge-sharing channels \parencite{diodato2018coagglom, steijn2022dynamics}, the shift in modern economies towards services may increase the share of input-output linkages that embed substantial tacit knowledge.

Third, the spatial units used in this paper, NUTS3 and NUTS4 regions, do not necessarily represent the most adequate spatial scales for all industries and all interactions. Future research could therefore use the exact geolocations of firms to construct coagglomeration patterns directly from the microgeography of firms.

Notwithstanding these limitations and open questions, we believe that our analysis advances our understanding of why industries coagglomerate, drawing attention to the importance of knowledge linkages along the value chain.

\section*{Acknowledgement} \label{sec:acknowledgement}

\noindent
Sándor Juhász’s work was supported by the European Union’s Marie Skłodowska-Curie Postdoctoral Fellowship Program (SUPPED, grant number 101062606). Zoltán Elekes and Virág Ilyés were supported by the Hungarian Scientiﬁc Research Fund project "Structure and robustness of regional supplier networks" (Grant No. OTKA FK-143064). Frank Neffke  acknowledges financial support from the Austrian Research Agency (FFG), project \#873927 (ESSENCSE). The authors would like to thank the Databank of HUN-REN Centre for Economic and Regional Studies for their support. This research paper is based on the value-added tax return data files of the Hungarian Central Statistical Office. The calculations and conclusions drawn from them are the sole intellectual property of the authors. The authors are grateful for the feedback of Mercedes Delgado, Max Nathan, Neave O'Clery, Cesar Hidalgo, Andrea Caragliu, Johannes Wachs and László Czaller. 

\singlespacing

\printbibliography

\clearpage
\section*{Supplementary information} \label{sec:si}

\setcounter{figure}{0}
\setcounter{table}{0}

\renewcommand{\thefigure}{SI\arabic{figure}}
\renewcommand{\thetable}{SI\arabic{table}}

\subsection*{S1 Description of the firm-level dataset} \label{sec:si_firm_data}

\begin{figure}[!ht]
\includegraphics[width=0.95\textwidth]{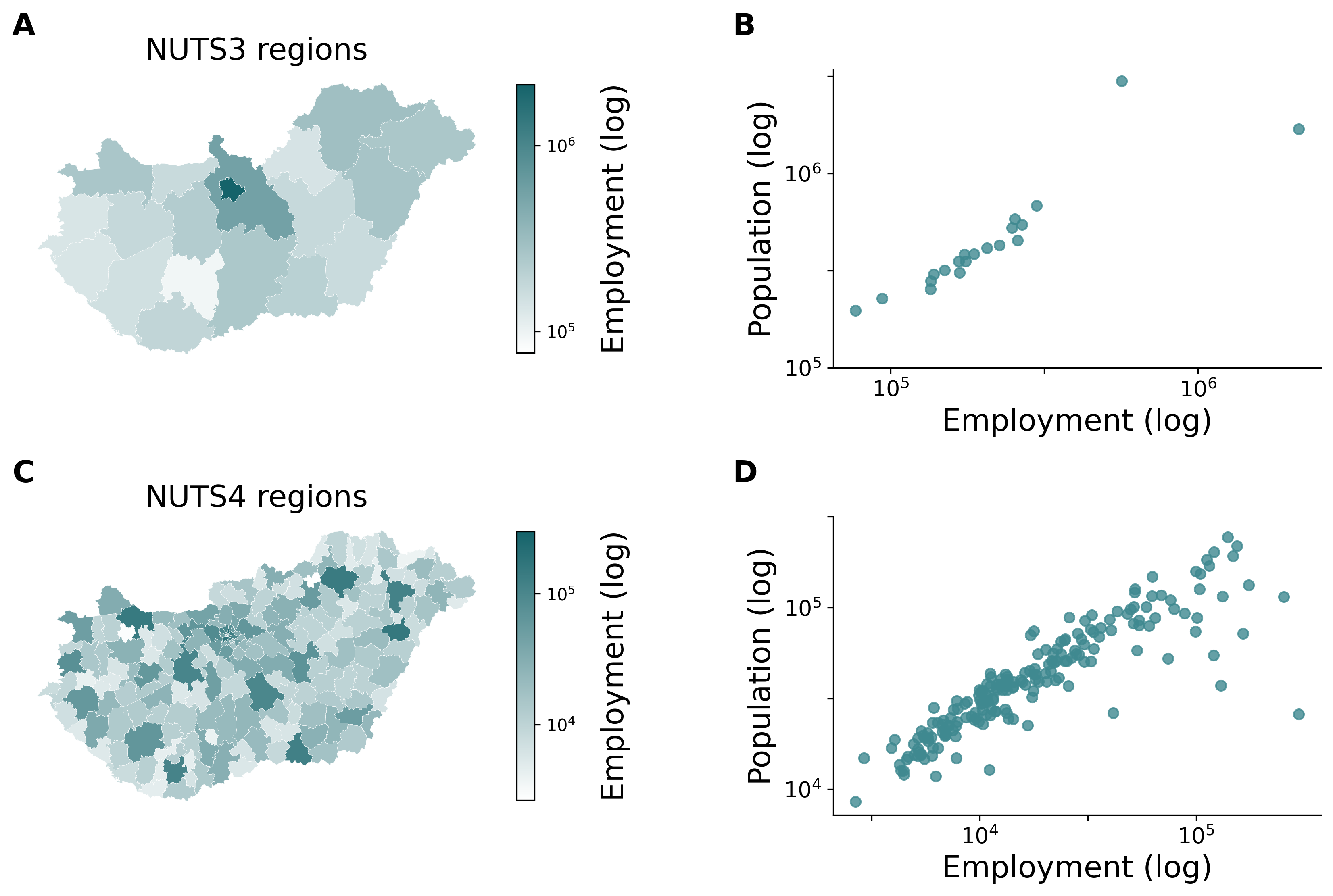}  
\caption{The spatial distribution of employment in our sample of firms used to calculate coagglomeration measures.
\textbf{(A)} Map of NUTS3 regions colored by employment in our sample.
\textbf{(B)} Correlation of employment in our sample and population at NUTS3 level.
\textbf{(C)} Map of NUTS4 regions colored by employment in our sample.
\textbf{(D)} Correlation of employment in our sample and population at NUTS4 level.}
\label{fig:si_map_emp_combined}
\end{figure}

The firm-level dataset we use to construct our sample of companies are made available by the Hungarian Central Statistical Office through the Databank of the HUN-REN Centre for Economic and Regional Studies\footnote{Databank of HUN-REN Centre for Economic and Regional Studies, https://adatbank.krtk.mta.hu/en/}. We focus on firms with at least 2 employees in 2017. Figure~\ref{fig:si_map_emp_combined} illustrates the spatial distribution of employment in our sample used to construct NUTS3 and NUTS4 level coagglomeration measures. Total employment in our sample is highly correlated with population at both NUTS3 and NUTS4 levels.

\begin{figure}[!ht]
\includegraphics[width=0.95\textwidth]{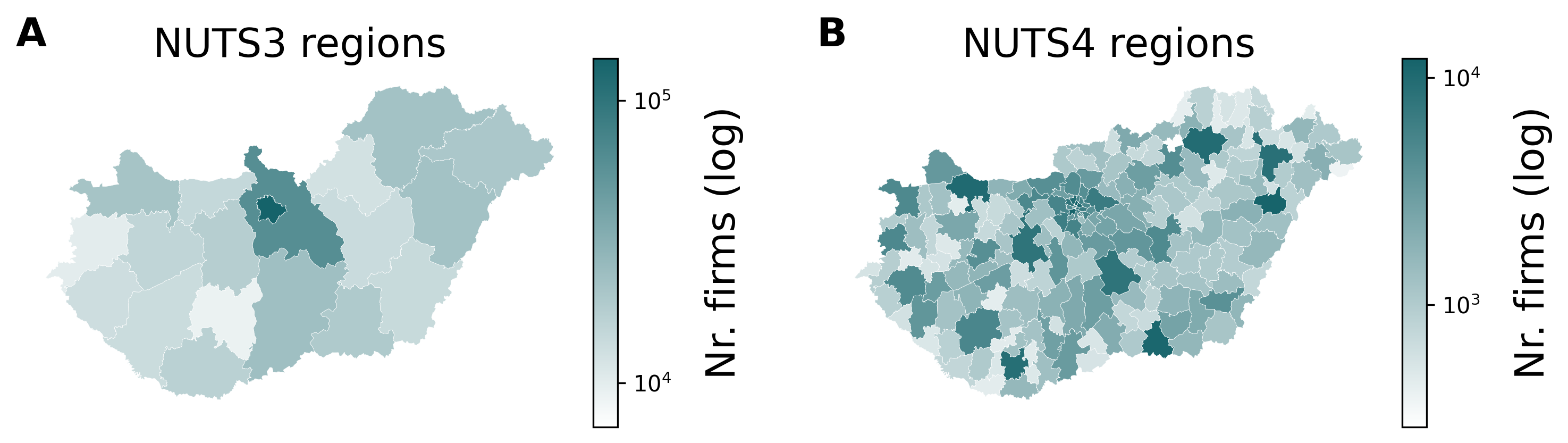}  
\caption{The spatial distribution of firms in our sample across NUTS3 \textbf{(A)} and NUTS4 \textbf{(B)} regions.}
\label{fig:si_map_firms}
\end{figure}

\begin{figure}[!ht]
\includegraphics[width=0.95\textwidth]{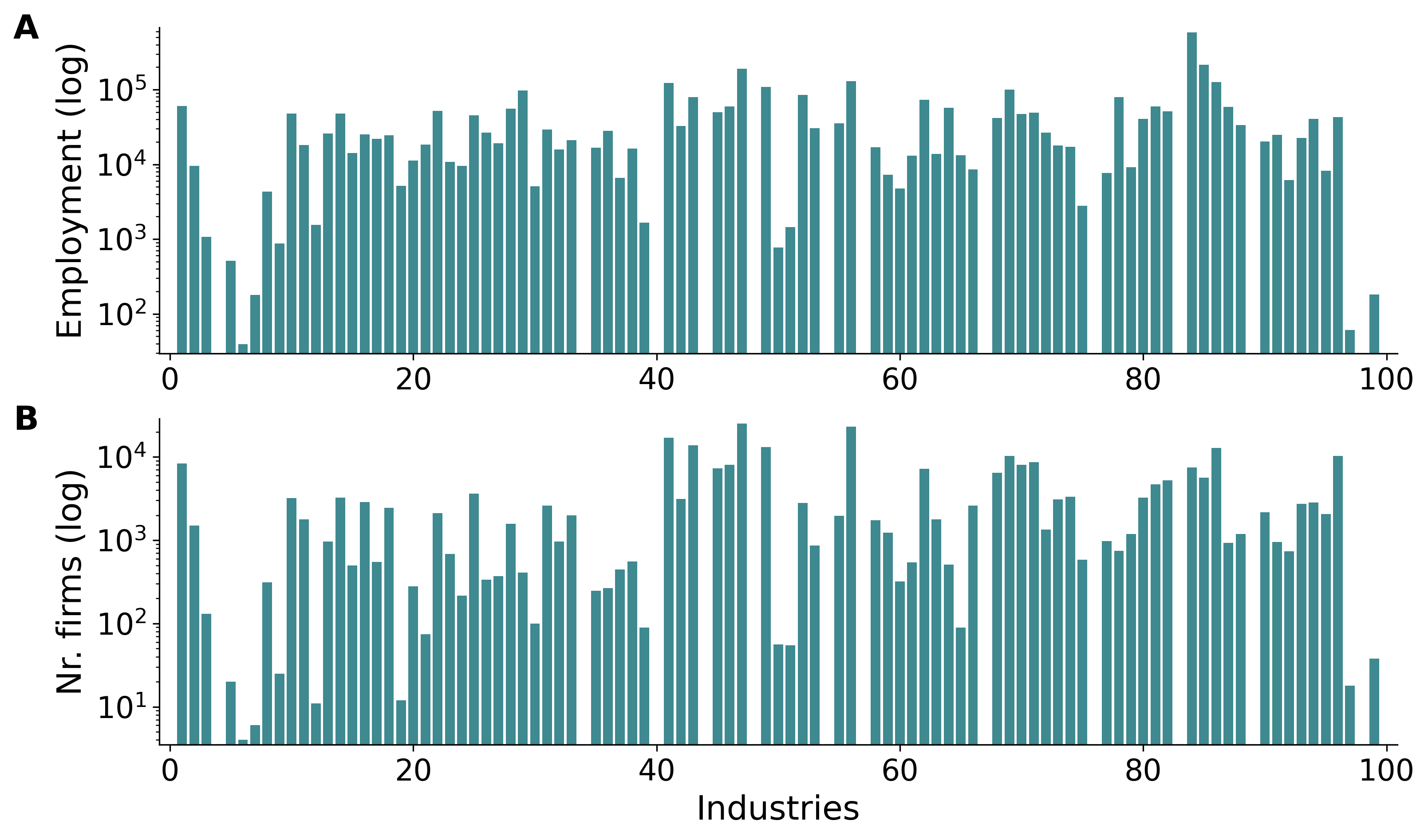}  
\caption{Distribution of employment \textbf{(A)} and firms \textbf{(B)} across industries. For visualization purposes we aggregated data to two-digit NACE codes.}
\label{fig:si_industries_distr}
\end{figure}

\clearpage
\subsection*{S2 Key variables and alternative measures} \label{sec:si_key_variables}

\begin{figure}[!ht]
\includegraphics[width=0.95\textwidth]{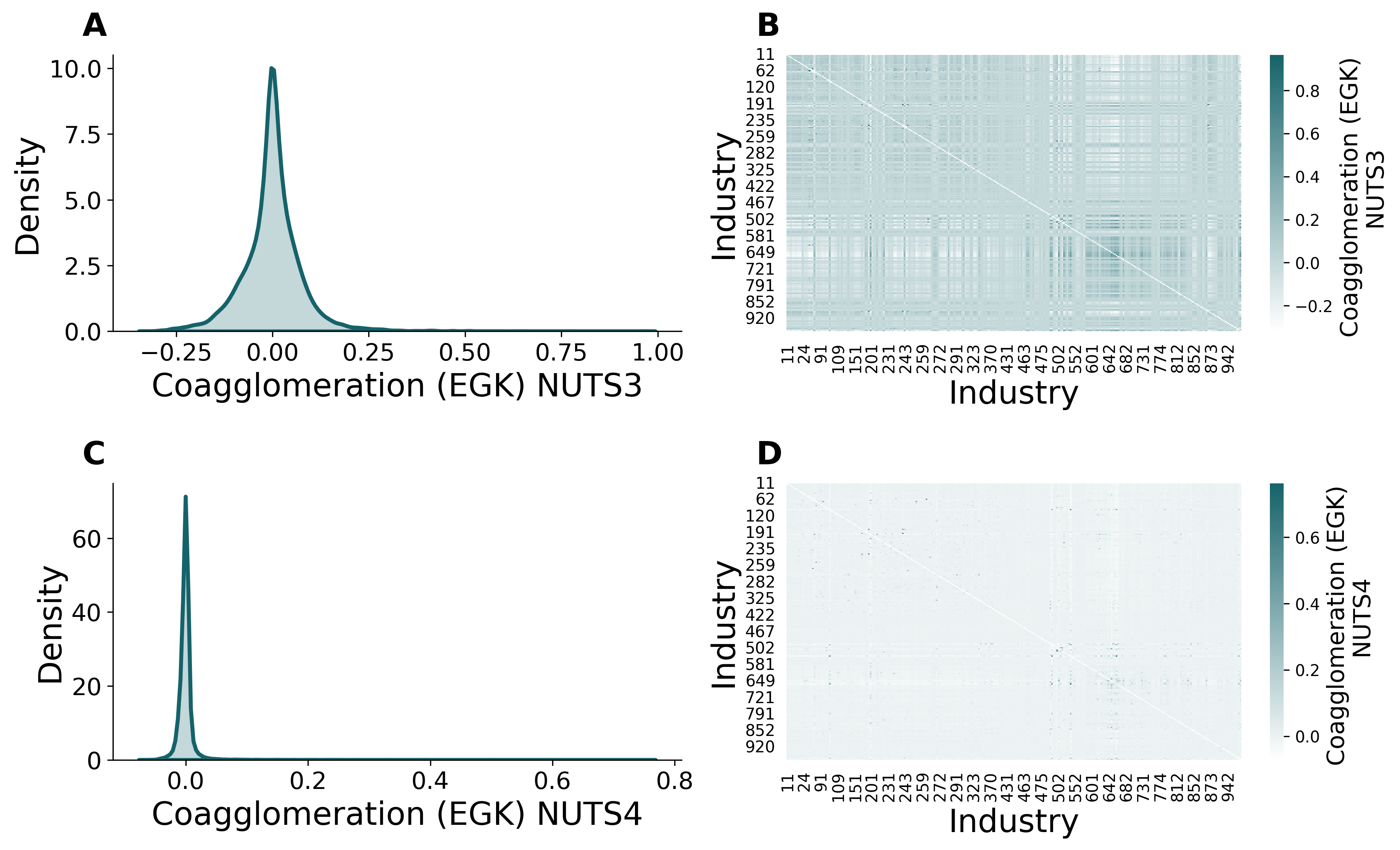}  
\caption{Distribution of EGK coagglomeration.
\textbf{(A)} Density of coagglomeration calculated at the NUTS3 level.
\textbf{(B)} Industry pair level illustration of coagglomeration values calculated at the NUTS3 level.
\textbf{(C)} Density of coagglomeration calculated at the NUTS4 level.
\textbf{(D)} Industry pair level illustration of coagglomeration values calculated at the NUTS4 level.}
\label{fig:si_egk_coagg_emp_dist}
\end{figure}

\begin{figure}[!ht]
\includegraphics[width=0.95\textwidth]{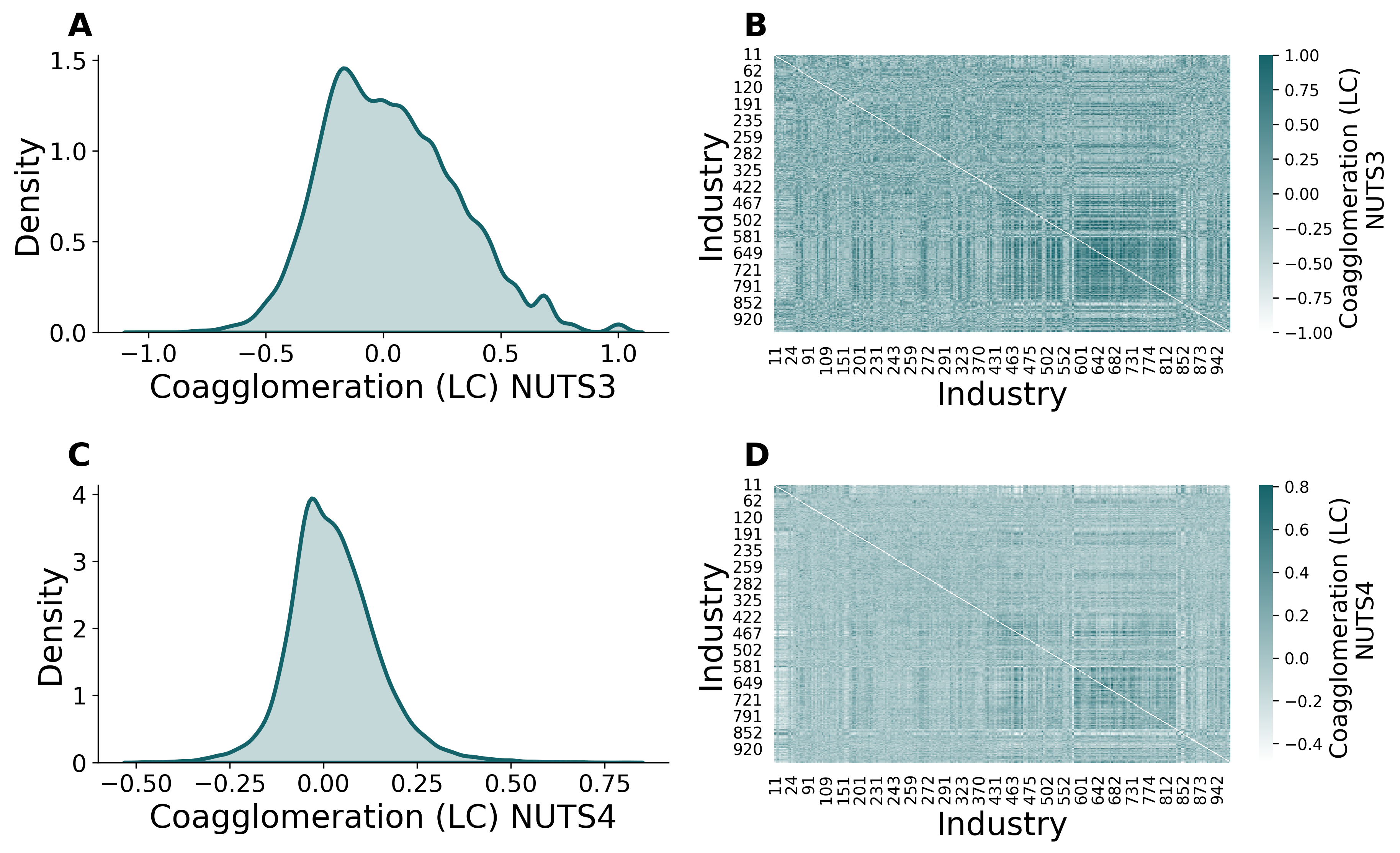}  
\caption{Distribution of Coagglomeration (LC) calculated on the basis of RCA 0/1 values.
\textbf{(A)} Density of coagglomeration calculated at the NUTS3 level.
\textbf{(B)} Industry pair level illustration of coagglomeration values calculated at the NUTS3 level.
\textbf{(C)} Density of coagglomeration calculated at the NUTS4 level.
\textbf{(D)} Industry pair level illustration of coagglomeration values calculated at the NUTS4 level.}
\label{fig:si_coagg_porter_emp_dist}
\end{figure}

To illustrate the robustness of our results we use two different dependent variables: \textit{EGK coagglomeration} and \textit{LC coagglomeration}. The top 10 industry pairs with the highest values for each of these indicators are reported in Table~\ref{tab:si_top10_egk} and in Table~\ref{tab:si_top10_lc}. The two measures rank two-digit industries differently. Figure~\ref{fig:si_egk_coagg_emp_dist} and Figure~\ref{fig:si_coagg_porter_emp_dist} present the distributions for both measures at different spatial scales in detail.

\begin{figure}[!ht]
\includegraphics[width=0.95\textwidth]{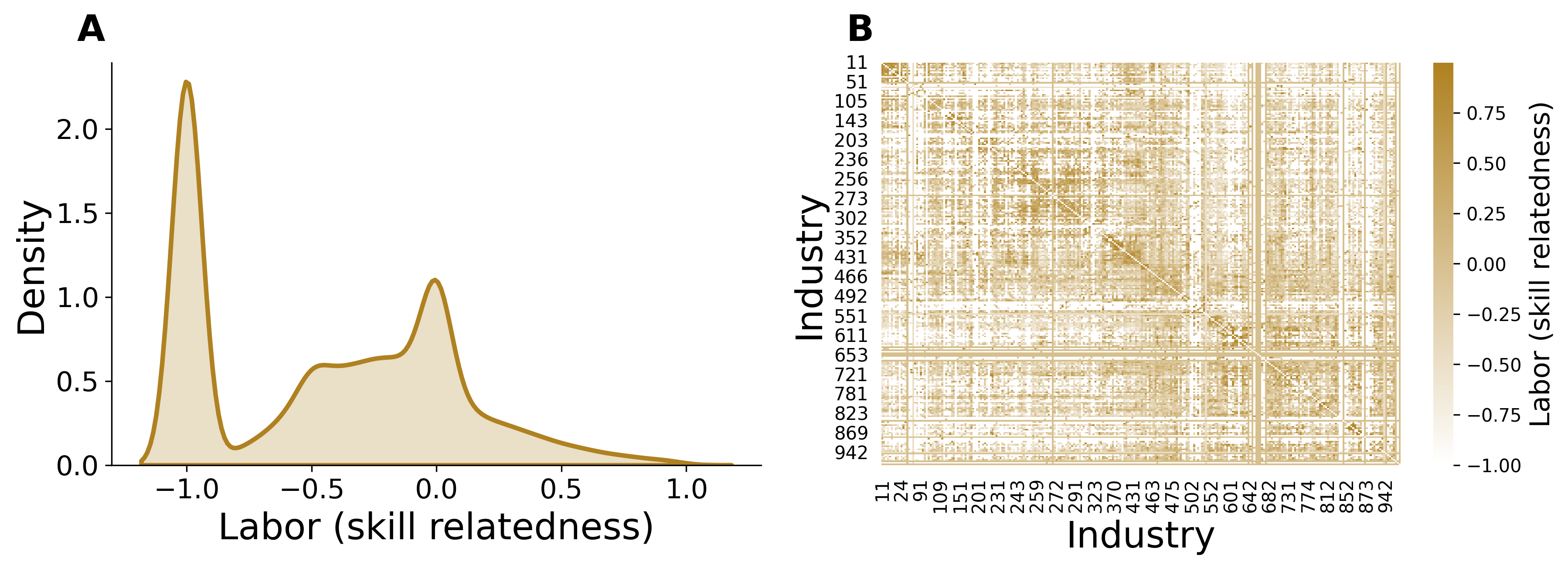}  
\caption{Distribution of our skill relatedness measure based on firm-to-firm labor flow in Hungary (2015-2017).
\textbf{(A)} Density of skill relatedness across all three-digit industries.
\textbf{(B)} Industry pair level illustration of skill relatedness values calculated at the three-digit level.}
\label{fig:si_sr_norm_dist}
\end{figure}

\begin{figure}[!ht]
\includegraphics[width=0.95\textwidth]{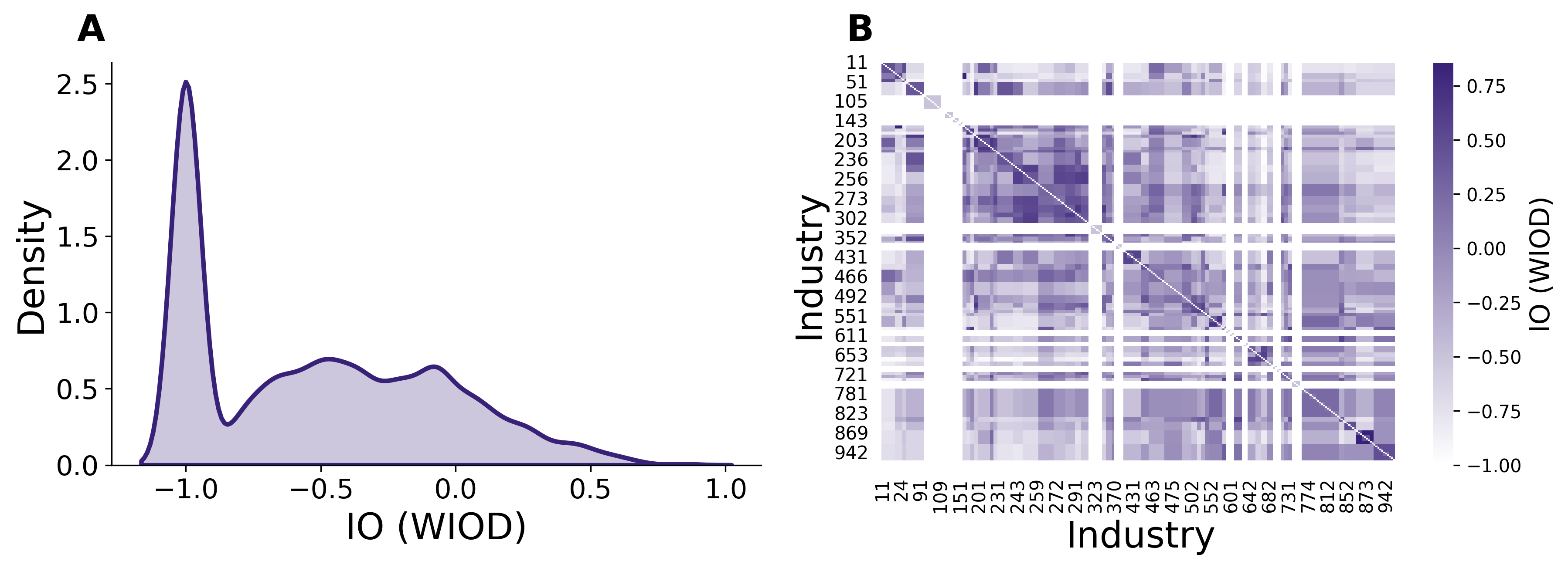}  
\caption{Distribution of our input-output similarity measure (IO) based on the World Input-Output Database (WIOD) for Hungary (2014).
\textbf{(A)} Density of our input-output similarity measure across all two-digit industries.
\textbf{(B)} Industry pair level illustration of input-output similarity values calculated at the two-digit level.}
\label{fig:si_io_wiot_dist}
\end{figure}

\begin{figure}[!ht]
\includegraphics[width=0.95\textwidth]{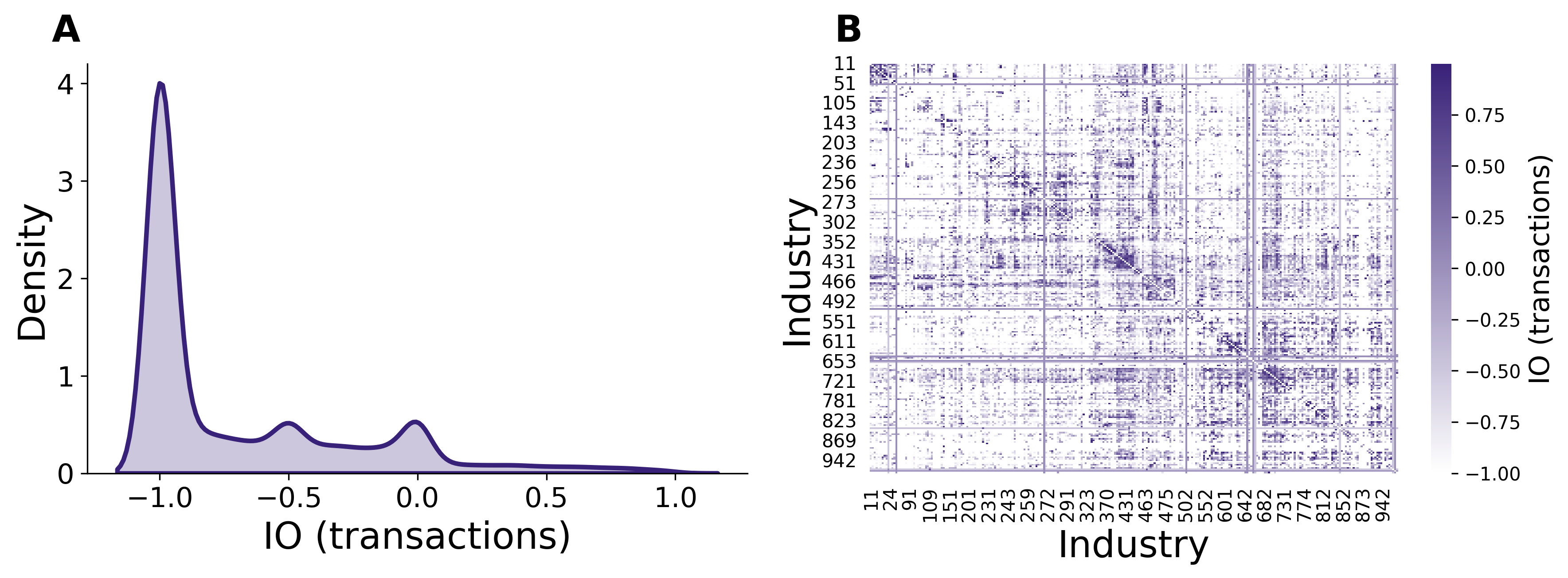}  
\caption{Distribution of our input-output similarity measure (IO) based on firm-to-firm transactions in Hungary (2015-2017).
\textbf{(A)} Density of our input-output similarity measure across all two-digit industries.
\textbf{(B)} Industry pair level illustration of input-output similarity values calculated at the three-digit level.}
\label{fig:si_io_transactions_dist}
\end{figure}

Figure~\ref{fig:si_sr_norm_dist} provides the description of the independent variable, skill relatedness, constructed from labor flows between industries in Hungary, 2015-2017. To measure input-output connections between industries we use two different datasets. Figure~\ref{fig:si_io_wiot_dist} presents the distributions for IO (WIOD), input-output connections of industries constructed from the World Input-Output Database (WIOD) for Hungary (2014). Note that this indicator is at the two-digit industry pair level. Figure~\ref{fig:si_io_transactions_dist} illustrates our alternative measure IO (transactions) at the three-digit industry pair level constructed from value added tax (VAT) reports.

\begin{figure}[!ht]
\includegraphics[width=0.95\textwidth]{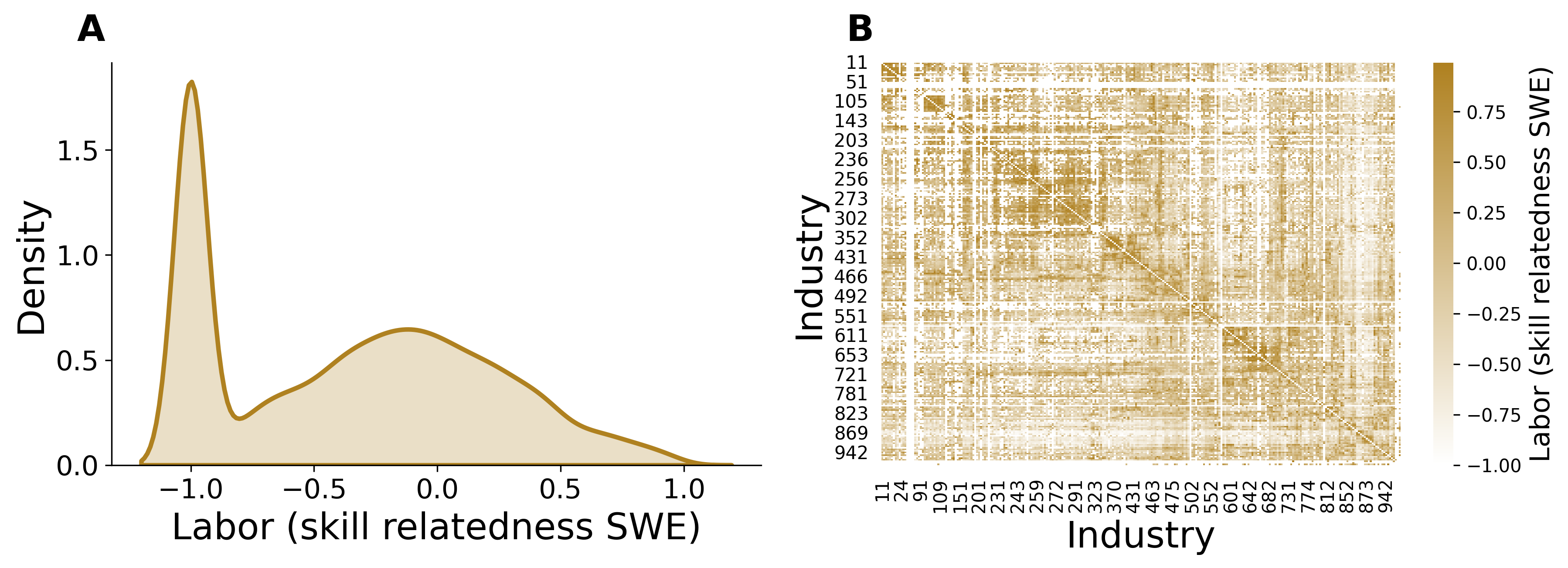}  
\caption{Distribution of our skill relatedness measure based on firm-to-firm labor flow in Sweden (2013-2019).
\textbf{(A)} Density of skill relatedness in Sweden across all three-digit industries.
\textbf{(B)} Industry pair level illustration of skill relatedness values calculated at the three-digit level for Sweden.}
\label{fig:si_swe_sr_norm_dist}
\end{figure}

We instrument Hungarian inter industry skill relatedness with a similarly constructed measure based on the Swedish economy. Specifically, we use a matched employer-employee data pooled from multiple Swedish registers, and made available by Statistics Sweden. This dataset covers the entire working age population and their workplaces on an annual basis for 2013-2019. Based on this dataset, we aggregate labor flows across three-digit industries (NACE Rev. 2 classification) over the sample period. From there we apply the same set of steps as with our main skill relatedness variable. Figure~\ref{fig:si_swe_sr_norm_dist} presents the distribution of the skill relatedness measure for Sweden (SWE).

\begin{figure}[!ht]
\includegraphics[width=0.95\textwidth]{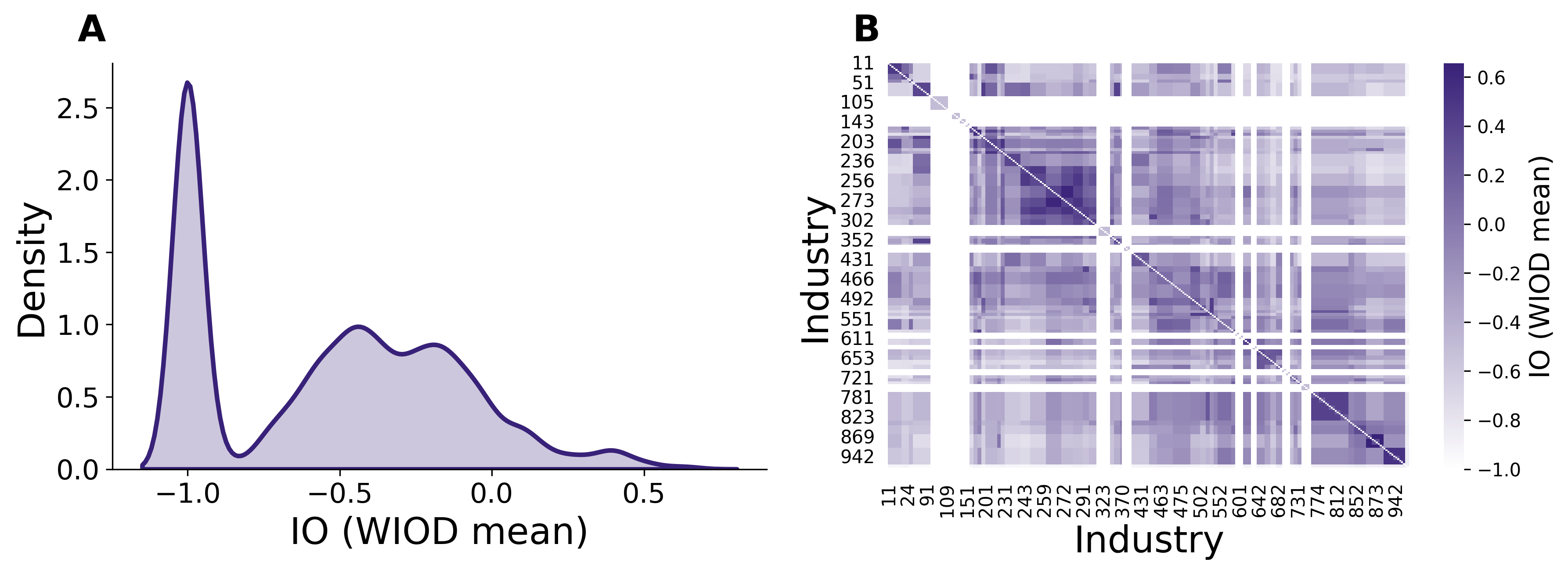}  
\caption{Distribution of our input-output similarity measure (IO) based on average flows across industries in the WIOD (excluding Hungary).
\textbf{(A)} Density of our input-output similarity measure across all three-digit industries.
\textbf{(B)} Industry pair level illustration of input-output similarity values calculated at the three-digit level.}
\label{fig:si_io_wiot_mean_dist}
\end{figure}

As an instrument for the Hungarian IO (WIOD) and IO (transactions) measures, we use the IO (WIOD mean) variable. It is constructed by excluding Hungary from the WIOD dataset and calculating the mean value for each input-output connection between industries. We use these mean values to create the same input-output measure as for IO (WIOD). Figure~\ref{fig:si_io_wiot_mean_dist} shows the distribution of the resulting indicator.

\begin{figure}[!t]
\includegraphics[width=0.95\textwidth]{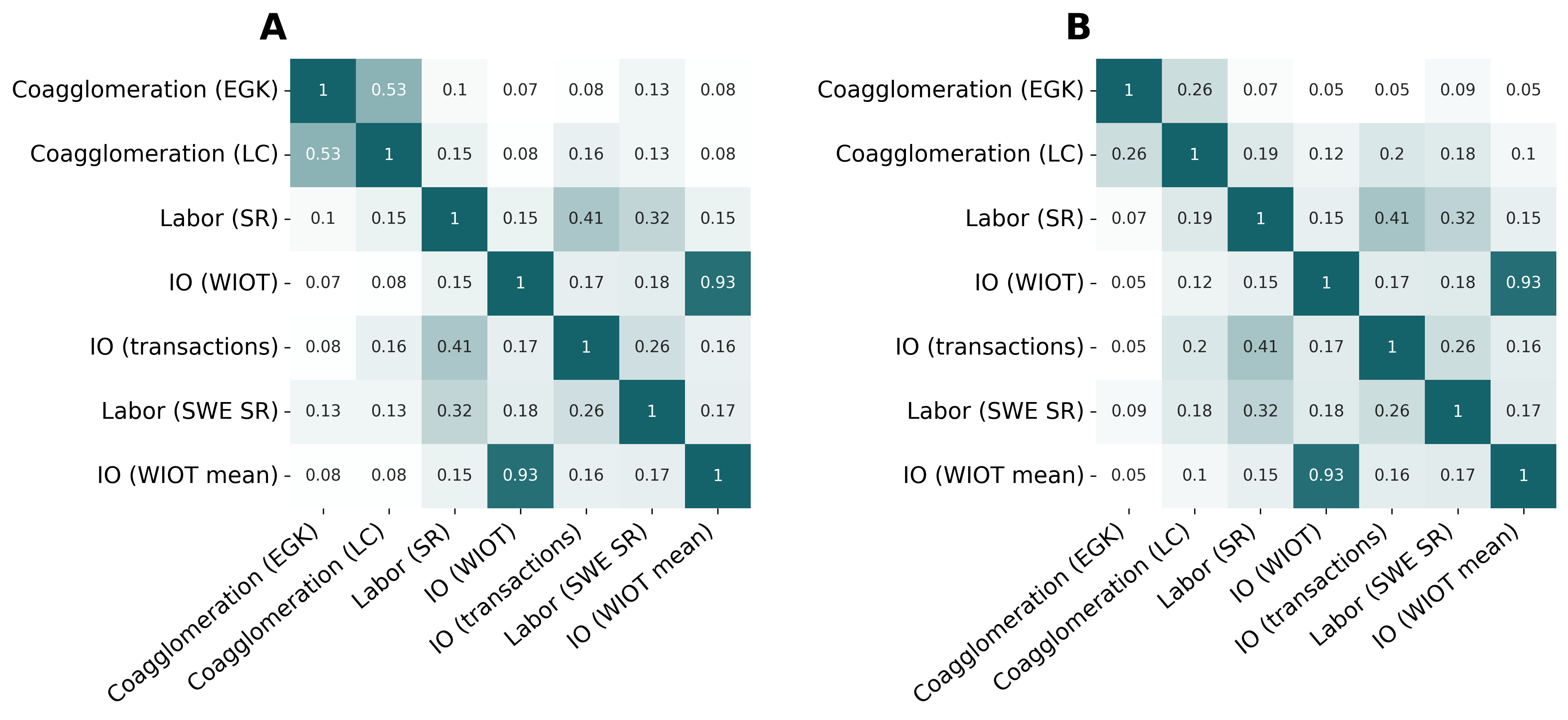}  
\caption{Correlation tables of key variables.
\textbf{(A)} Correlations with coagglomeration measures constructed at NUTS3 scale.
\textbf{(B)} Correlations with coagglomeration measures constructed at NUTS4 scale.}
\label{fig:si_corr_mat}
\end{figure}

Figure~\ref{fig:si_corr_mat}A presents the correlation matrix for all our key variables with coagglomeration measures constructed at NUTS3 level, while Figure~\ref{fig:si_corr_mat}B shows correlation with coagglomeration measures constructed at NUTS4 level.

\begin{table}[!htbp]
\begin{minipage}{\linewidth}
\centering
\fontsize{10}{12}\selectfont
\captionsetup{singlelinecheck=off, skip=0.ex}
\caption{Top 10 most colocated indsutries} 
\label{tab:si_top10_egk} 
\begin{tabular}{@{\extracolsep{1pt}}llll}
\\[-1.8ex]\hline 
\\[-1.8ex]\multicolumn{3}{l}{EGK coagglomeration, NUTS4} \\
\\[-1.8ex]\hline 
\\[-1.8ex]Industry i & Industry j & Value \\
\\[-2.5ex]\hline 
\\[-1.8ex]Manuf. of coke oven products & Manuf. of basic iron and steel & 0.762 \\
\\[-1.8ex]Postal activities & Reinsurance & 0.735 \\
\\[-1.8ex]Freight rail transport & Postal activities & 0.619 \\
\\[-1.8ex]Extraction of crude petroleum & Manuf. of computers and equip. & 0.499 \\
\\[-1.8ex]Freight rail transport & Reinsurance & 0.495 \\
\\[-1.8ex]Manuf. of vegetable and animal oils & Postal activities & 0.493 \\
\\[-1.8ex]Manuf. of pulp, paper and paperboard & Manuf. of coke oven products & 0.445 \\
\\[-1.8ex]Transport via pipeline & Inland freight water transport  & 0.443 \\
\\[-1.8ex]Sea and coastal freight water transport & Trusts, funds and sim. financial entities & 0.408 \\
\\[-1.8ex]Manuf. of vegetable and animal oils & Reinsurance & 0.394 \\
\hline \\[-1.8ex] 
\end{tabular}
\caption*{\textbf{Note:} Some names are shown in abbreviated form. \textit{EGK coagglomeration} is calculated at NUTS4 level.}
\end{minipage}
\end{table}

\begin{table}[!htbp]
\begin{minipage}{\linewidth}
\centering
\fontsize{10}{12}\selectfont
\captionsetup{singlelinecheck=off, skip=0.ex}
\caption{Top 10 most colocated indsutries}
\label{tab:si_top10_lc} 
\begin{tabular}{@{\extracolsep{1pt}}llll}
\\[-1.8ex]\hline 
\\[-1.8ex]\multicolumn{3}{l}{LC coagglomeration, NUTS4} \\
\\[-1.8ex]\hline 
\\[-1.8ex]Industry i & Industry j & Value \\
\\[-2.5ex]\hline 
\\[-1.8ex]Other financial serv., except insurance & Fund management & 0.807 \\
\\[-1.8ex]Insurance & Fund management & 0.734 \\
\\[-1.8ex]Wholesale of household goods & Non-specialised wholesale trade & 0.701 \\
\\[-1.8ex]Motion picture, video and television & Computer programming, consultancy & 0.697 \\
\\[-1.8ex]Wholesale of household goods & Management consultancy & 0.681 \\
\\[-1.8ex]Fund management & Act. of extraterritorial org. & 0.680 \\
\\[-1.8ex]Wholesale of infocom. equiptment & Non-spec. wholesale trade  & 0.677 \\
\\[-1.8ex]Computer programming, consultancy & Data processing, hosting & 0.676 \\
\\[-1.8ex]Software publishing  & Computer programming, consultancy & 0.675 \\
\\[-1.8ex]Other passenger land transport & Compulsory social security activities & 0.670 \\
\hline \\[-1.8ex] 
\end{tabular}
\caption*{\textbf{Note:} Some names are shown in abbreviated form. \textit{LC coagglomeration} is calculated at NUTS4 level.}
\end{minipage}
\end{table}

\clearpage
\subsection*{S3 Univariate OLS estimations for labor and input-output channels} \label{sec:ols_univar}

Univariate regression models to assess the relationship between coagglomeration and labor flows (Table ~\ref{tab:ols_univar_labor}) and between coagglomeration and input-output relationships (Table ~\ref{tab:ols_univar_io}) across various settings.

\begin{table}[!htbp]
\begin{minipage}{\linewidth}
\centering
\fontsize{10}{12}\selectfont
\captionsetup{singlelinecheck=off, skip=0.ex}
\caption{Labor channel through OLS univariate regressions} 
\label{tab:ols_univar_labor}
\begin{tabular}{@{\extracolsep{1pt}}lcccc} 
\\[-1.8ex]\hline 
\\[-1.8ex] & \multicolumn{2}{c}{Coagglomeration (EGK)} & \multicolumn{2}{c}{Coagglomeration (LC)} \\  
\\[-2.5ex] & NUTS3 & NUTS4 & NUTS3 & NUTS4\\ 
\\[-2.5ex] & (1) & (2) & (3) & (4)\\ 
\hline \\[-1.8ex]
Labor (SR) & 0.104$^{***}$ & 0.067$^{***}$ & 0.151$^{***}$ & 0.189$^{***}$ \\
& (0.019) & (0.011) & (0.018) & (0.023)\\
\hline \\[-1.8ex] 
Observations & 35,778 & 35,778 & 35,778 & 35,778 \\ 
R$^{2}$ & 0.011 & 0.004 & 0.023 & 0.036\\ 
Adjusted R$^{2}$ & 0.011 & 0.004 & 0.023 & 0.036\\ 
\hline \\[-1.8ex]
\end{tabular}
\caption*{\textbf{Note:} Clustered (industry$_i$ and industry$_j$) standard errors in parentheses.  Significance codes: ***: p$<$0.01, **: p$<$0.05, *: p$<$0.1.}
\end{minipage}
\end{table}

\begin{table}[!htbp]
\begin{minipage}{\linewidth}
\centering
\fontsize{10}{12}\selectfont
\captionsetup{singlelinecheck=off, skip=0.ex}
\caption{Input-output channel through OLS univariate regressions} 
\label{tab:ols_univar_io}
\begin{tabular}{@{\extracolsep{1pt}}lcccccccc} 
\\[-1.8ex]\hline 
\\[-1.8ex] & \multicolumn{4}{c}{Coagglomeration (EGK)} & \multicolumn{4}{c}{Coagglomeration (LC)} \\  
\\[-2.5ex] & NUTS3 & NUTS4 & NUTS3 & NUTS4 & NUTS3 & NUTS4 & NUTS3 & NUTS4\\ 
\\[-2.5ex] & (1) & (2) & (3) & (4) & (5) & (6) & (7) & (8)\\ 
\hline \\[-1.8ex]
IO (WIOD) & 0.070$^{***}$ & 0.050$^{***}$ & & & 0.085$^{***}$ & 0.115$^{***}$ & & \\
& (0.018) & (0.011) & & & (0.020) & (0.021) & & \\ 
IO (transactions) & & & 0.084$^{***}$ & 0.049$^{***}$ & & & 0.155$^{***}$ & 0.205$^{***}$\\
& & & (0.023) & (0.015) & & & (0.018) & (0.023)\\
\hline \\[-1.8ex] 
Observations & 35,778 & 35,778 & 35,778 & 35,778 & 35,778 & 35,778 & 35,778 & 35,778 \\ 
R$^{2}$ & 0.005 & 0.002 & 0.007 & 0.002 & 0.007 & 0.013 & 0.024 & 0.042\\ 
Adjusted R$^{2}$ & 0.005 & 0.002 & 0.007 & 0.002 & 0.007 & 0.013 & 0.024 & 0.042\\ 
\hline \\[-1.8ex]
\end{tabular}
\caption*{\textbf{Note:} Clustered (industry$_i$ and industry$_j$) standard errors in parentheses.  Significance codes: ***: p$<$0.01, **: p$<$0.05, *: p$<$0.1.}
\end{minipage}
\end{table}

\clearpage
\subsection*{S4 Stepwise instrumental variable estimations} \label{sec:stepwise_instruments}

Table~\ref{tab:si_stepwise_iv_labor} presents first and second stage instrumental variable estimations focusing on the labor market channel only. Table~\ref{tab:si_stepwise_iv_wiod} shows the same two-stage instrumental variable estimations using IO (WIOD) as the proxy for the input-output channel. Table~\ref{tab:si_stepwise_iv_trans} presents the same two-stage estimation focusing on IO (transactions) only.

\begin{table}[!ht]
\begin{minipage}{\linewidth}
\centering
\fontsize{10}{12}\selectfont
\captionsetup{singlelinecheck=off, skip=0.ex}
\caption{First and second stage instrumental variable regressions for labor similarity} 
\label{tab:si_stepwise_iv_labor} 
\begin{tabular}{@{\extracolsep{1pt}}lccccc}
\hline \\[-1.8ex] 
\\[-1.8ex] & First stage & \multicolumn{4}{c}{Second stage} \\ 
\\[-1.8ex] & Labor (SR HUN) & \multicolumn{2}{c}{Coagglomeration (EGK)} & \multicolumn{2}{c}{Coagglomeration (LC)} \\ 
\\[-2.5ex]  &  & NUTS3 & NUTS4 & NUTS3 & NUTS4\\ 
\\[-2.5ex]  & (1) & (2) & (3) & (4) & (5)\\ 
\hline \\[-1.8ex] 
Labor (SR SWE) & 0.319$^{***}$ &  &  &  &  \\ 
& (0.030) &  &  &  &  \\ 
Labor pred &  & 0.403$^{***}$ & 0.285$^{***}$ & 0.417$^{***}$ & 0.575$^{***}$ \\ 
&  & (0.060) & (0.030) & (0.060) & (0.070) \\ 
\hline \\[-1.8ex] 
Observations & 35,778 & 35,778 & 35,778 & 35,778 & 35,778 \\ 
R$^{2}$ & 0.102 & 0.017 & 0.008 & 0.018 & 0.034 \\ 
Adjusted R$^{2}$ & 0.102 & 0.017 & 0.008 & 0.018 & 0.034 \\ 
\hline \\[-1.8ex]
\end{tabular} 
\caption*{\textbf{Note:} Clustered (industry$_i$ and industry$_j$) standard errors in parentheses.  Significance codes: ***: p$<$0.01, **: p$<$0.05, *: p$<$0.1.}
\end{minipage}
\end{table}

\begin{table}[!ht]
\begin{minipage}{\linewidth}
\centering
\fontsize{10}{12}\selectfont
\captionsetup{singlelinecheck=off, skip=0.ex}
\caption{First and second stage instrumental variable regressions for input-output connections based on WIOD} 
\label{tab:si_stepwise_iv_wiod} 
\begin{tabular}{@{\extracolsep{1pt}}lccccc}
\hline \\[-1.8ex] 
\\[-1.8ex] & First stage & \multicolumn{4}{c}{Second stage} \\ 
\\[-1.8ex] & IO (WIOD HUN) & \multicolumn{2}{c}{Coagglomeration (EGK)} & \multicolumn{2}{c}{Coagglomeration (LC)} \\ 
\\[-2.5ex]  &  & NUTS3 & NUTS4 & NUTS3 & NUTS4\\ 
\\[-2.5ex]  & (1) & (2) & (3) & (4) & (5)\\ 
\hline \\[-1.8ex] 
IO (WIOD mean) & 0.929$^{***}$ &  &  &  &  \\ 
& (0.009) &  &  &  &  \\ 
IO pred &  & 0.083$^{***}$ & 0.054$^{***}$ & 0.083$^{***}$ & 0.106$^{***}$ \\ 
&  & (0.020) & (0.012) & (0.023) & (0.023) \\ 
\hline \\[-1.8ex] 
Observations & 35,778 & 35,778 & 35,778 & 35,778 & 35,778 \\ 
R$^{2}$ & 0.863 & 0.006 & 0.003 & 0.006 & 0.010 \\ 
Adjusted R$^{2}$ & 0.863 & 0.006 & 0.002 & 0.006 & 0.010 \\ 
\hline \\[-1.8ex]
\end{tabular}
\caption*{\textbf{Note:} Clustered (industry$_i$ and industry$_j$) standard errors in parentheses.  Significance codes: ***: p$<$0.01, **: p$<$0.05, *: p$<$0.1.}
\end{minipage}
\end{table}

\begin{table}[!ht]
\begin{minipage}{\linewidth}
\centering
\fontsize{10}{12}\selectfont
\captionsetup{singlelinecheck=off, skip=0.ex}
\caption{Stepwise instrumental variable regressions for input-output connections based on transactions} 
\label{tab:si_stepwise_iv_trans} 
\begin{tabular}{@{\extracolsep{1pt}}lccccc}
\hline \\[-1.8ex] 
\\[-1.8ex] & First stage & \multicolumn{4}{c}{Second stage} \\ 
\\[-1.8ex] & IO (transactions) & \multicolumn{2}{c}{Coagglomeration (EGK)} & \multicolumn{2}{c}{Coagglomeration (LC)} \\ 
\\[-2.5ex]  &  & NUTS3 & NUTS4 & NUTS3 & NUTS4\\ 
\\[-2.5ex]  & (1) & (2) & (3) & (4) & (5)\\ 
\hline \\[-1.8ex] 
IO (WIOD mean) & 0.161$^{***}$ &  &  &  &  \\ 
& (0.026) &  &  &  &  \\ 
IO pred &  & 0.475$^{***}$ & 0.310$^{***}$ & 0.479$^{***}$ & 0.609$^{***}$ \\ 
&  & (0.115) & (0.067) & (0.130) & (0.132) \\ 
\hline \\[-1.8ex] 
Observations & 35,778 & 35,778 & 35,778 & 35,778 & 35,778 \\ 
R$^{2}$ & 0.026 & 0.006 & 0.003 & 0.006 & 0.010 \\ 
Adjusted R$^{2}$ & 0.026 & 0.006 & 0.002 & 0.006 & 0.010 \\ 
\hline \\[-1.8ex]
\end{tabular} 
\caption*{\textbf{Note:} Clustered (industry$_i$ and industry$_j$) standard errors in parentheses.  Significance codes: ***: p$<$0.01, **: p$<$0.05, *: p$<$0.1.}
\end{minipage}
\end{table}

\clearpage
\subsection*{S5 Alternative instruments for input-output connections} \label{sec:si_alt_instruments}

Table~\ref{tab:si_iv_us_supply} presents an alternative instrumental variable regression where we instrument IO (WIOD) and IO (transactions) data from US supply tables (2017) at the three-digit NACE code level.
Table~\ref{tab:si_iv_cze_wiod} presents the same estimations using only the WIOD table of the Czech Republic as an instrument. Our results are stable across these alternative instrumental variable settings.

\begin{table}[!htbp]
\begin{minipage}{\linewidth}
\centering
\fontsize{10}{12}\selectfont
\captionsetup{singlelinecheck=off, skip=0.ex}
\caption{Instrumental variable regressions using US supply tables as instruments for IO connections} 
\label{tab:si_iv_us_supply}
\begin{tabular}{@{\extracolsep{1pt}}lcccccccc} 
\\[-1.8ex]\hline 
\\[-1.8ex] & \multicolumn{4}{c}{Coagglomeration (EGK)} & \multicolumn{4}{c}{Coagglomeration (LC)} \\  
\\[-2.5ex] & NUTS3 & NUTS4 & NUTS3 & NUTS4 & NUTS3 & NUTS4 & NUTS3 & NUTS4\\ 
\\[-2.5ex] & (1) & (2) & (3) & (4) & (5) & (6) & (7) & (8)\\ 
\hline \\[-1.8ex]
IO (WIOD) & 0.372$^{***}$ & 0.247$^{***}$ & & & 0.448$^{***}$ & 0.638$^{***}$ & & \\
& (0.088) & (0.052) & & & (0.087) & (0.121) & & \\
IO (transactions) & & & 0.339$^{***}$ & 0.225$^{***}$ & & & 0.407$^{***}$ & 0.581$^{***}$\\
& & & (0.063) & (0.037) & & & (0.057) & (0.072)\\
\hline \\[-1.8ex] 
Observations & 29,890 & 29,890 & 29,890 & 29,890 & 29,890 & 29,890 & 29,890 & 29,890 \\ 
R$^{2}$ & -0.144 & -0.065 & -0.071 & -0.033 & -0.139 & -0.272 & -0.043 & -0.062 \\ 
Adjusted R$^{2}$ & -0.144 & -0.065 & -0.071 & -0.033 & -0.139 & -0.272 & -0.043 & -0.062 \\
KP F-statistic & 33.249 & 33.249 & 83.337 & 83.337 & 33.249 & 33.249 & 83.337 & 83.337 \\ 
\hline \\[-1.8ex]
\end{tabular}
\caption*{\textbf{Note:} Clustered (industry$_i$ and industry$_j$) standard errors in parentheses. Significance codes: ***: p$<$0.01, **: p$<$0.05, *: p$<$0.1.}
\end{minipage}
\end{table}

\begin{table}[!htbp]
\begin{minipage}{\linewidth}
\centering
\fontsize{10}{12}\selectfont
\captionsetup{singlelinecheck=off, skip=0.ex}
\caption{Instrumental variable regressions using the WIOD for Czech Republic as instrument for IO connections} 
\label{tab:si_iv_cze_wiod}
\begin{tabular}{@{\extracolsep{1pt}}lcccccccc} 
\\[-1.8ex]\hline 
\\[-1.8ex] & \multicolumn{4}{c}{Coagglomeration (EGK)} & \multicolumn{4}{c}{Coagglomeration (LC)} \\  
\\[-2.5ex] & NUTS3 & NUTS4 & NUTS3 & NUTS4 & NUTS3 & NUTS4 & NUTS3 & NUTS4\\ 
\\[-2.5ex] & (1) & (2) & (3) & (4) & (5) & (6) & (7) & (8)\\ 
\hline \\[-1.8ex]
IO (WIOD) & 0.082$^{***}$ & 0.058$^{***}$ & & & 0.063$^{***}$ & 0.074$^{***}$ & & \\
& (0.022) & (0.014) & & & (0.022) & (0.022) & & \\ 
IO (transactions) & & & 0.479$^{***}$ & 0.343$^{***}$ & & & 0.370$^{***}$ & 0.438$^{***}$ \\
& & & (0.141) & (0.094) & & & (0.124) & (0.132)\\
\hline \\[-1.8ex] 
Observations & 35,778 & 35,778 & 35,778 & 35,778 & 35,778 & 35,778 & 35,778 & 35,778 \\ 
R$^{2}$ & -0.070 & -0.039 & -0.123 & -0.073 & -0.047 & -0.118 & -0.031 & -0.226 \\ 
Adjusted R$^{2}$ & -0.070 & -0.039 & -0.123 & -0.074 & -0.047 & -0.118 & -0.031 & -0.226 \\ 
KP F-statistic & 2570.495 & 2570.495 & 31.631 & 31.631 & 2570.495 & 2570.495 & 31.631 & 31.631 \\ 
\hline \\[-1.8ex]
\end{tabular}
\caption*{\textbf{Note:} Clustered (industry$_i$ and industry$_j$) standard errors in parentheses. Significance codes: ***: p$<$0.01, **: p$<$0.05, *: p$<$0.1.}
\end{minipage}
\end{table}

\clearpage
\subsection*{S6 Results for manufacturing and services} \label{sec:si_manuf_services}

Our work studies coagglomeration of all sectors of the economy. However, we enable comparisons to previous studies and provide robustness check by reproducing our main regressions for manufacturing and service sectors separately. This is done by spliting the rows of the industry-industry matrix used to measure coagglomeration into a manufacturing and a services section. How industries coagglomerate with any other industry is studied within these subsets as suggested by \textcite{diodato2018coagglom}. In other words, we model the coagglomeration patterns of a manufacturing industry (or of a service) with all other industries.

Our main results seem to hold when focusing on manufacturing, but for the service sector we can only successfully instrument the labor channel. This result is in line with previous findings of \textcite{diodato2018coagglom}, which highlight the more prominent role of labor channels for the coagglomeration of service sector industries.

\begin{table}[!htbp]
\begin{minipage}{\linewidth}
\centering
\fontsize{10}{12}\selectfont
\captionsetup{singlelinecheck=off, skip=0.ex}
\caption{OLS multivariate regressions only for manufacturing} 
\label{tab:si_manufacturing_ols} 
\begin{tabular}{@{\extracolsep{1pt}}lcccccccc} 
\\[-1.8ex]\hline 
\\[-1.8ex] & \multicolumn{4}{c}{Coagglomeration (EGK)} & \multicolumn{4}{c}{Coagglomeration (LC)} \\  
\\[-2.5ex] & NUTS3 & NUTS4 & NUTS3 & NUTS4 & NUTS3 & NUTS4 & NUTS3 & NUTS4\\ 
\\[-2.5ex] & (1) & (2) & (3) & (4) & (5) & (6) & (7) & (8)\\ 
\hline \\[-1.8ex]
Labor (SR) & 0.099$^{***}$ & 0.123$^{***}$ & 0.112$^{***}$ & 0.126$^{***}$ & 0.076$^{***}$ & 0.079$^{***}$ & 0.075$^{***}$ & 0.067$^{***}$\\
& (0.031) & (0.022) & (0.029) & (0.027) & (0.017) & (0.016) & (0.018) & (0.015)\\ 
IO (WIOD) & 0.006 & 0.042$^{**}$ & & & 0.080$^{***}$ & 0.088$^{***}$ & & \\
& (0.032) & (0.018) & & & (0.020) & (0.018) & & \\   
IO (transactions) & & & -0.029 & 0.018 & & & 0.051$^{***}$ & 0.086$^{***}$\\
& & & (0.022) & (0.031) & & & (0.017) & (0.017)\\ 
\hline \\[-1.8ex] 
Observations & 4,536 & 4,536 & 4,536 & 4,536 & 4,536 & 4,536 & 4,536 & 4,536 \\ 
R$^2$ & 0.019 & 0.018 & 0.020 & 0.017 & 0.032 & 0.053 & 0.020 & 0.039\\
Adjusted R$^2$ & 0.019 & 0.018 & 0.020 & 0.016 & 0.032 & 0.053 & 0.020 & 0.039\\
\hline \\[-1.8ex]
\end{tabular} 
\caption*{\textbf{Note:} Clustered (industry$_i$ and industry$_j$) standard errors in parentheses.  Significance codes: ***: p$<$0.01, **: p$<$0.05, *: p$<$0.1.}
\end{minipage}
\end{table}

\begin{table}[!htbp]
\begin{minipage}{\linewidth}
\centering
\fontsize{10}{12}\selectfont
\captionsetup{singlelinecheck=off, skip=0.ex}
\caption{Labor channel through instrumental variable univariate regressions for manufacturing only} 
\label{tab:si_manufacturing_iv_labor}
\begin{tabular}{@{\extracolsep{1pt}}lcccc} 
\\[-1.8ex]\hline 
\\[-1.8ex] & \multicolumn{2}{c}{Coagglomeration (EGK)} & \multicolumn{2}{c}{Coagglomeration (LC)} \\  
\\[-2.5ex] & NUTS3 & NUTS4 & NUTS3 & NUTS4\\ 
\\[-2.5ex] & (1) & (2) & (3) & (4)\\ 
\hline \\[-1.8ex]
Labor (SR) & 0.107$^{**}$ & 0.134$^{***}$ & 0.186$^{***}$ & 0.175$^{***}$ \\
& (0.052) & (0.029) & (0.047) & (0.044) \\  
\hline \\[-1.8ex] 
Observations & 4,536 & 4,536 & 4,536 & 4,536 \\ 
R$^{2}$ & 0.019 & 0.017 & 0.002 & 0.012 \\
Adjusted R$^{2}$ & 0.019 & 0.016 & 0.002 & 0.012 \\
KP F-statistic & 109.122 & 109.122 & 109.122 & 109.122 \\ 
\hline \\[-1.8ex]
\end{tabular}
\caption*{\textbf{Note:} Clustered (industry$_i$ and industry$_j$) standard errors in parentheses.  Significance codes: ***: p$<$0.01, **: p$<$0.05, *: p$<$0.1.}
\end{minipage}
\end{table}

\begin{table}[!htbp]
\begin{minipage}{\linewidth}
\centering
\fontsize{10}{12}\selectfont
\captionsetup{singlelinecheck=off, skip=0.ex}
\caption{Input-output channel through instrumental variable univariate regressions for manufacturing only} 
\label{tab:si_manufacturing_iv_io}
\begin{tabular}{@{\extracolsep{1pt}}lcccccccc}
\\[-1.8ex]\hline 
\\[-1.8ex] & \multicolumn{4}{c}{Coagglomeration (EGK)} & \multicolumn{4}{c}{Coagglomeration (LC)} \\  
\\[-2.5ex] & NUTS3 & NUTS4 & NUTS3 & NUTS4 & NUTS3 & NUTS4 & NUTS3 & NUTS4\\ 
\\[-2.5ex] & (1) & (2) & (3) & (4) & (5) & (6) & (7) & (8)\\ 
\hline \\[-1.8ex]
IO (WIOD) & 0.025 & 0.064$^{***}$ & & & 0.097$^{***}$ & 0.116$^{***}$ & & \\
& (0.030) & (0.021) & & & (0.022) & (0.019) & & \\ 
IO (transactions) & & & 0.093 & 0.240$^{***}$ & & & 0.361$^{***}$ & 0.432$^{***}$ \\
& & & (0.112) & (0.080) & & & (0.081) & (0.076) \\ 
\hline \\[-1.8ex] 
Observations & 4,536 & 4,536 & 4,536 & 4,536 & 4,536 & 4,536 & 4,536 & 4,536 \\ 
R$^{2}$ & 0.002 & 0.005 & -0.005 & -0.012 & 0.022 & 0.036 & -0.094 & -0.170 \\ 
Adjusted R$^{2}$ & 0.001 & 0.005 & -0.005 & -0.013 & 0.022 & 0.036 & -0.094 & -0.171\\
KP F-statistic & 4375.016 & 4375.016 & 50.764 & 50.764 & 4375.016 & 4375.016 & 50.764 & 50.764\\
\hline \\[-1.8ex]
\end{tabular}
\caption*{\textbf{Note:} Clustered (industry$_i$ and industry$_j$)  standard errors in parentheses.  Significance codes: ***: p$<$0.01, **: p$<$0.05, *: p$<$0.1.}
\end{minipage}
\end{table}

\begin{table}[!htbp]
\begin{minipage}{\linewidth}
\centering
\fontsize{10}{12}\selectfont
\captionsetup{singlelinecheck=off, skip=0.ex}
\caption{OLS multivariate regressions only for services} 
\label{tab:si_services_ols} 
\begin{tabular}{@{\extracolsep{1pt}}lcccccccc} 
\\[-1.8ex]\hline 
\\[-1.8ex] & \multicolumn{4}{c}{Coagglomeration (EGK)} & \multicolumn{4}{c}{Coagglomeration (LC)} \\  
\\[-2.5ex] & NUTS3 & NUTS4 & NUTS3 & NUTS4 & NUTS3 & NUTS4 & NUTS3 & NUTS4\\ 
\\[-2.5ex] & (1) & (2) & (3) & (4) & (5) & (6) & (7) & (8)\\ 
\hline \\[-1.8ex]
Labor (SR) & 0.178$^{***}$ & 0.105$^{***}$ & 0.101$^{**}$ & 0.068$^{***}$ & 0.232$^{***}$ & 0.300$^{***}$ & 0.151$^{***}$ & 0.197$^{***}$ \\
& (0.049) & (0.022) & (0.048) & (0.017) & (0.051) & (0.058) & (0.048) & (0.055)\\ 
IO (WIOD) & -0.028 & 0.045 & & & -0.082$^{**}$ & -0.079$^{*}$ & & \\
& (0.039) & (0.031) & & & (0.037) & (0.041) & & \\  
IO (transactions) & & & 0.176$^{***}$ & 0.088$^{***}$ & & & 0.185$^{***}$ & 0.235$^{***}$ \\
& & & (0.046) & (0.025) & & & (0.040) & (0.041) \\  
\hline \\[-1.8ex] 
Observations & 5,886 & 5,886 & 5,886 & 5,886 & 5,886 & 5,886 & 5,886 & 5,886 \\ 
R$^2$ & 0.024 & 0.006 & 0.048 & 0.009 & 0.043 & 0.061 & 0.065 & 0.095\\
Adjusted R$^2$ & 0.024 & 0.006 & 0.048 & 0.009 & 0.043 & 0.061 & 0.065 & 0.095\\
\hline \\[-1.8ex]
\end{tabular} 
\caption*{\textbf{Note:} Clustered (industry$_i$ and industry$_j$) standard errors in parentheses.  Significance codes: ***: p$<$0.01, **: p$<$0.05, *: p$<$0.1.}
\end{minipage}
\end{table}

\begin{table}[!htbp]
\begin{minipage}{\linewidth}
\centering
\fontsize{10}{12}\selectfont
\captionsetup{singlelinecheck=off, skip=0.ex}
\caption{Labor channel through instrumental variable univariate regressions for services only} 
\label{tab:si_services_iv_labor}
\begin{tabular}{@{\extracolsep{1pt}}lcccc} 
\\[-1.8ex]\hline 
\\[-1.8ex] & \multicolumn{2}{c}{Coagglomeration (EGK)} & \multicolumn{2}{c}{Coagglomeration (LC)} \\  
\\[-2.5ex] & NUTS3 & NUTS4 & NUTS3 & NUTS4\\ 
\\[-2.5ex] & (1) & (2) & (3) & (4)\\ 
\hline \\[-1.8ex]
Labor (SR) & 1.170$^{***}$ & 0.598$^{***}$ & 1.288$^{***}$ & 1.784$^{***}$\\   
& (0.265) & (0.182) & (0.270) & (0.372)\\  
\hline \\[-1.8ex] 
Observations & 5,886 & 5,886 & 5,886 & 5,886 \\ 
R$^{2}$ & -0.714 & -0.108 & -0.768 & -1.36\\
Adjusted R$^{2}$& -0.714 & -0.108 & -0.768 & -1.36\\
KP F-statistic & 31.371 & 31.371 & 31.371 & 31.371\\ 
\hline \\[-1.8ex]
\end{tabular}
\caption*{\textbf{Note:} Clustered (industry$_i$ and industry$_j$) standard errors in parentheses.  Significance codes: ***: p$<$0.01, **: p$<$0.05, *: p$<$0.1.}
\end{minipage}
\end{table}

\begin{table}[!htbp]
\begin{minipage}{\linewidth}
\centering
\fontsize{10}{12}\selectfont
\captionsetup{singlelinecheck=off, skip=0.ex}
\caption{Input-output channel through instrumental variable univariate regressions for services only} 
\label{tab:si_services_iv_io}
\begin{tabular}{@{\extracolsep{1pt}}lcccccccc} 
\\[-1.8ex]\hline 
\\[-1.8ex] & \multicolumn{4}{c}{Coagglomeration (EGK)} & \multicolumn{4}{c}{Coagglomeration (LC)} \\  
\\[-2.5ex] & NUTS3 & NUTS4 & NUTS3 & NUTS4 & NUTS3 & NUTS4 & NUTS3 & NUTS4\\ 
\\[-2.5ex] & (1) & (2) & (3) & (4) & (5) & (6) & (7) & (8)\\ 
\hline \\[-1.8ex]
IO (WIOD) & -0.047 & 0.040 & & & -0.106$^{**}$ & -0.116$^{**}$ & & \\   
& (0.043) & (0.034) & & & (0.044) & (0.050) & & \\ 
IO (transactions) & & & 13.3 & -11.4 & & & 30.1 & 33.3\\
& & & (190.9) & (170.8) & & & (437.7) & (481.4)\\ 
\hline \\[-1.8ex] 
Observations & 5,886 & 5,886 & 5,886 & 5,886 & 5,886 & 5,886 & 5,886 & 5,886 \\ 
R$^{2}$ & 0.000 & 0.001 & -160.3 & -76.8 & 0.004 & 0.002 & -807.5 & -875.1\\ 
Adjusted R$^{2}$ & -0.000 & 0.001 & -160.3 & -76.9 & 0.004 & 0.002 & -807.6 & -875.2\\
KP F-statistic & 3388.765 & 3388.765 & 0.005 & 0.005 & 3388.765 & 3388.765 & 0.005 & 0.005\\
\hline \\[-1.8ex]
\end{tabular}
\caption*{\textbf{Note:} Clustered (industry$_i$ and industry$_j$) standard errors in parentheses.  Significance codes: ***: p$<$0.01, **: p$<$0.05, *: p$<$0.1.}
\end{minipage}
\end{table}

\clearpage
\subsection*{S7 Geographic restrictions for robustness checks} \label{sec:si_geo_limitations}

To increase the validity of our results, we run regressions on restricted samples. Table~\ref{tab:si_budapest_excluded_ols} presents OLS and while Table~\ref{tab:si_budapest_excluded_iv} presents IV models on a sample that excludes the 23 NUTS4 regions of Budapest, the capital city of Hungary, from the construction of our dependent variables. Table~\ref{tab:si_singleplant_ols} presents OLS and Table~\ref{tab:si_singleplant_iv} presents IV models in a similar fashion on a sample that only involves firms with a single plant. In nearly all settings our main results remain the same.

\begin{table}[!htbp]
\begin{minipage}{\linewidth}
\centering
\fontsize{10}{12}\selectfont
\captionsetup{singlelinecheck=off, skip=0.ex}
\caption{OLS multivariate regressions excluding Budapest from the sample} 
\label{tab:si_budapest_excluded_ols} 
\begin{tabular}{@{\extracolsep{1pt}}lcccc} 
\\[-1.8ex]\hline 
\\[-1.8ex] & \multicolumn{2}{c}{Coagglomeration (EGK)} & \multicolumn{2}{c}{Coagglomeration (LC)} \\  
\\[-2.5ex] & (1) & (2) & (3) & (4)\\ 
\hline \\[-1.8ex]
Labor (SR) & 0.076$^{***}$ & 0.070$^{***}$ & 0.150$^{***}$ & 0.104$^{***}$\\
& (0.012) & (0.012) & (0.022) & (0.020)\\ 
IO (WIOD) & 0.008 & & 0.052$^{**}$ & \\
& (0.008) & & (0.021) & \\   
IO (transactions) & & 0.017$^{*}$ & & 0.132$^{***}$\\
& & (0.009) & & (0.017)\\  
\hline \\[-1.8ex] 
Observations & 34,980 & 34,980 & 34,980 & 34,980\\ 
R$^2$ & 0.006 & 0.006 & 0.028 & 0.039\\
Adjusted R$^2$ & 0.006 & 0.006 & 0.028 & 0.039\\
\hline \\[-1.8ex]
\end{tabular} 
\caption*{\textbf{Note:} Clustered (industry$_i$ and industry$_j$)  standard errors in parentheses.  Significance codes: ***: p$<$0.01, **: p$<$0.05, *: p$<$0.1.}
\end{minipage}
\end{table}

\begin{table}[!htbp]
\begin{minipage}{\linewidth}
\centering
\fontsize{10}{12}\selectfont
\captionsetup{singlelinecheck=off, skip=0.ex}
\caption{Instrumental variable univariate regressions excluding Budapest from the sample} 
\label{tab:si_budapest_excluded_iv} 
\begin{tabular}{@{\extracolsep{1pt}}lcccccc} 
\\[-1.8ex]\hline 
\\[-1.8ex] & \multicolumn{3}{c}{Coagglomeration (EGK)} & \multicolumn{3}{c}{Coagglomeration (LC)} \\  
\\[-2.5ex] & (1) & (2) & (3) & (4) & (5) & (6)\\ 
\hline \\[-1.8ex]
Labor (SR) & 0.185$^{***}$ & & & 0.423$^{***}$ & & \\
& (0.028) & & & (0.062) & & \\ 
IO (WIOD) & & 0.024$^{***}$ & & & 0.062$^{***}$ & \\
& & (0.008) & & & (0.022) & \\   
IO (transactions) & & & 0.141$^{***}$ & & & 0.368$^{***}$\\
& & & (0.050) & & & (0.125)\\ 
\hline \\[-1.8ex] 
Observations & 34,980 & 34,980 & 34,980 & 34,980 & 34,980 & 34,980\\ 
R$^2$ & -0.006 & 0.0004 & -0.007 & -0.045 & 0.005 & -0.006\\
Adjusted R$^2$ & -0.006 & 0.0004 & -0.007 & -0.045 & 0.005 & -0.006\\
KP F-statistic & 131.455 & 11000 & 35.132 & 131.455 & 11000 & 35.132\\
\hline \\[-1.8ex]
\end{tabular} 
\caption*{\textbf{Note:} Clustered (industry$_i$ and industry$_j$) standard errors in parentheses.  Significance codes: ***: p$<$0.01, **: p$<$0.05, *: p$<$0.1.}
\end{minipage}
\end{table}

\begin{table}[!htbp]
\begin{minipage}{\linewidth}
\centering
\fontsize{10}{12}\selectfont
\captionsetup{singlelinecheck=off, skip=0.ex}
\caption{OLS multivariate regressions based on single plant companies only} 
\label{tab:si_singleplant_ols} 
\begin{tabular}{@{\extracolsep{1pt}}lcccc} 
\\[-1.8ex]\hline 
\\[-1.8ex] & \multicolumn{2}{c}{Coagglomeration (EGK)} & \multicolumn{2}{c}{Coagglomeration (LC)} \\  
\\[-2.5ex] & (1) & (2) & (3) & (4)\\ 
\hline \\[-1.8ex]
Labor (SR) & 0.099$^{***}$ & 0.093$^{***}$ & 0.107$^{***}$ & 0.082$^{***}$\\
& (0.010) & (0.013) & (0.021) & (0.017)\\ 
IO (WIOD) & 0.012 & & 0.086$^{***}$ & \\
& (0.009) & & (0.023) & \\   
IO (transactions) & & 0.021 & & 0.100$^{***}$\\
& & (0.014) & & (0.018)\\
\hline \\[-1.8ex] 
Observations & 30,876 & 30,876 & 30,876 & 30,876\\ 
R$^2$ & 0.011 & 0.011 & 0.023 & 0.024\\
Adjusted R$^2$ & 0.011 & 0.011 & 0.022 & 0.024\\
\hline \\[-1.8ex]
\end{tabular} 
\caption*{\textbf{Note:} Clustered (industry$_i$ and industry$_j$) standard errors in parentheses.  Significance codes: ***: p$<$0.01, **: p$<$0.05, *: p$<$0.1.}
\end{minipage}
\end{table}

\begin{table}[!htbp]
\begin{minipage}{\linewidth}
\centering
\fontsize{10}{12}\selectfont
\captionsetup{singlelinecheck=off, skip=0.ex}
\caption{Instrumental variable univariate regressions based on single plant companies only} 
\label{tab:si_singleplant_iv} 
\begin{tabular}{@{\extracolsep{1pt}}lcccccc} 
\\[-1.8ex]\hline 
\\[-1.8ex] & \multicolumn{3}{c}{Coagglomeration (EGK)} & \multicolumn{3}{c}{Coagglomeration (LC)} \\  
\\[-2.5ex] & (1) & (2) & (3) & (4) & (5) & (6)\\ 
\hline \\[-1.8ex]
Labor (SR) & 0.183$^{***}$ & & & 0.270$^{***}$ & & \\
& (0.023) & & & (0.042) & & \\
IO (WIOD) & & 0.031$^{***}$ & & & 0.098$^{***}$ & \\
& & (0.009) & & & (0.024) & \\  
IO (transactions) & & & 0.177$^{***}$ & & & 0.554$^{***}$\\
& & & (0.055) & & & (0.152)\\ 
\hline \\[-1.8ex] 
Observations & 30,876 & 30,876 & 30,876 & 30,876 & 30,876 & 30,876\\ 
R$^2$ & 0.004 & 0.001 & -0.011 & -0.007 & 0.011 & -0.161\\
Adjusted R$^2$ & 0.004 & 0.001 & -0.011 & -0.007 & 0.011 & -0.161\\
KP F-statistic & 371.583 & 11000 & 36.578 & 371.583 & 11000 & 36.578\\
\hline \\[-1.8ex]
\end{tabular} 
\caption*{\textbf{Note:} Clustered (industry$_i$ and industry$_j$) standard errors in parentheses.  Significance codes: ***: p$<$0.01, **: p$<$0.05, *: p$<$0.1.}
\end{minipage}
\end{table}

\clearpage
\subsection*{S8 Multivariate instrumental variable regressions} \label{sec:si_multiver_ivs}

Using the same structure as our multivariate OLS models in the main text, we present multivariate instrumental variable regressions in Table~\ref{tab:iv_multivar_cse} with clustered standard errors and in Table~\ref{tab:iv_multivar_robust} with robust standard errors. The Kleibergen-Paap F-statistic suggests that the majority of these models suffer from weak instruments.

\begin{table}[!htbp]
\begin{minipage}{\linewidth}
\centering
\fontsize{10}{12}\selectfont
\captionsetup{singlelinecheck=off, skip=0.ex}
\caption{Instrumental variable multivariate regressions} 
\label{tab:iv_multivar_cse}
\begin{tabular}{@{\extracolsep{1pt}}lcccccccc} 
\\[-1.8ex]\hline 
\\[-1.8ex] & \multicolumn{4}{c}{Coagglomeration (EGK)} & \multicolumn{4}{c}{Coagglomeration (LC)} \\  
\\[-2.5ex] & NUTS3 & NUTS4 & NUTS3 & NUTS4 & NUTS3 & NUTS4 & NUTS3 & NUTS4\\ 
\\[-2.5ex] & (1) & (2) & (3) & (4) & (5) & (6) & (7) & (8)\\ 
\hline \\[-1.8ex]
Labor (SR) & 0.391$^{***}$ & 0.279$^{***}$ & 0.083 & 0.135 & 0.407$^{***}$ & 0.566$^{***}$ & 0.125 & 0.322 \\ 
& (0.075) & (0.045) & (0.343) & (0.207) & (0.071) & (0.090) & (0.412) & (0.497)\\ 
IO (WIOD) & 0.021 & 0.010 &  &  & 0.019 & 0.016 &  & \\ 
& (0.021) & (0.013) & & & (0.027) & (0.032) & & \\
IO (transactions) &  &  & 0.400 & 0.187 &  &  & 0.365 & 0.317 \\ 
& & & (0.410) & (0.247) & & & (0.488) & (0.574)\\
\hline \\[-1.8ex] 
Observations & 35,778 & 35,778 & 35,778 & 35,778 & 35,778 & 35,778 & 35,778 & 35,778 \\ 
R$^{2}$ & -0.071 & -0.041 & -0.109 & -0.037 & -0.042 & -0.106 & -0.035 & -0.036 \\ 
Adjusted R$^{2}$ & -0.071 & -0.041 & -0.109 & -0.037 & -0.042 & -0.106 & -0.035 & -0.036 \\ 
KP F-statistic & 51.432 & 51.432 & 2.505 & 2.505 & 51.432 & 51.432 & 2.505 & 2.505 \\ 
\hline \\[-1.8ex]
\end{tabular}
\caption*{\textbf{Note:} Clustered (industry$_i$ and industry$_j$) standard errors in parentheses.  Significance codes: ***: p$<$0.01, **: p$<$0.05, *: p$<$0.1.}
\end{minipage}
\end{table}

\begin{table}[!htbp]
\begin{minipage}{\linewidth}
\centering
\fontsize{10}{12}\selectfont
\captionsetup{singlelinecheck=off, skip=0.ex}
\caption{Instrumental variable multivariate regressions} 
\label{tab:iv_multivar_robust}
\begin{tabular}{@{\extracolsep{1pt}}lcccccccc} 
\\[-1.8ex]\hline 
\\[-1.8ex] & \multicolumn{4}{c}{Coagglomeration (EGK)} & \multicolumn{4}{c}{Coagglomeration (LC)} \\  
\\[-2.5ex] & NUTS3 & NUTS4 & NUTS3 & NUTS4 & NUTS3 & NUTS4 & NUTS3 & NUTS4\\ 
\\[-2.5ex] & (1) & (2) & (3) & (4) & (5) & (6) & (7) & (8)\\ 
\hline \\[-1.8ex]
Labor (SR) & 0.391$^{***}$ & 0.279$^{***}$ & 0.083 & 0.135 & 0.407$^{***}$ & 0.566$^{***}$ & 0.125 & 0.322$^{***}$ \\ 
& (0.020) & (0.022) & (0.112) & (0.120) & (0.019) & (0.019) & (0.111) & (0.113)\\
IO (WIOD) & 0.021$^{***}$ & 0.010 &  &  & 0.019$^{***}$ & 0.016$^{***}$ &  & \\ 
& (0.007) & (0.007) & & & (0.007) & (0.007) & & \\
IO (transactions) &  &  & 0.400$^{***}$ & 0.187 &  &  & 0.365$^{***}$ & 0.317$^{***}$ \\ 
& & & (0.130) & (0.139) & & & (0.131) & (0.132)\\
\hline \\[-1.8ex] 
Observations & 35,778 & 35,778 & 35,778 & 35,778 & 35,778 & 35,778 & 35,778 & 35,778 \\ 
R$^{2}$ & -0.071 & -0.041 & -0.109 & -0.037 & -0.042 & -0.106 & -0.035 & -0.035 \\ 
Adjusted R$^{2}$ & -0.071 & -0.041 & -0.109 & -0.037 & -0.042 & -0.106 & -0.035 & -0.036 \\ 
KP F-statistic & 1438.047 & 1438.047 & 30.43 & 30.43 & 1438.047 & 1438.047 & 30.43 & 30.43 \\ 
\hline \\[-1.8ex]
\end{tabular}
\caption*{\textbf{Note:} Robust standard errors in parentheses. Significance codes: ***: p$<$0.01, **: p$<$0.05, *: p$<$0.1.}
\end{minipage}
\end{table}

\clearpage
\subsection*{S9 Alternative specifications for our models with interactions}
\label{sec:si_interactions_nuts3}

Table~\ref{tab:si_interactions_nuts3} presents OLS regressions with interactions using NUTS3 level dependent variables, while Table~\ref{tab:si_interactions_budapestexcl} presents results based on NUTS4 level dependent variables in case we exclude regions of Budapest. For further robustness checks in Table~\ref{tab:si_interactions_manuf} and Table~\ref{tab:si_interactions_serv} we present the same models as in the main text for manufacturing and services only.

\begin{table}[!htbp]
\begin{minipage}{\linewidth}
\centering
\fontsize{10}{12}\selectfont
\captionsetup{singlelinecheck=off, skip=0.ex}
\caption{OLS multivariate regressions with interaction effects (NUTS3 level)} 
\label{tab:si_interactions_nuts3} 
\begin{tabular}{@{\extracolsep{0.75pt}}lcccccccc} 
\\[-1.8ex]\hline 
\\[-1.8ex] & \multicolumn{4}{c}{Coagglomeration (EGK)} & \multicolumn{4}{c}{Coagglomeration (LC)} \\ 
\\[-2.5ex] & NUTS3 & NUTS3 & NUTS3 & NUTS3 & NUTS3 & NUTS3 & NUTS3 & NUTS3\\ 
\\[-2.5ex] & (1) & (2) & (3) & (4) & (5) & (6) & (7) & (8) \\ 
\hline \\[-1.8ex] 
Labor (SR) & 0.096$^{***}$ & 0.072$^{***}$ & 0.084$^{***}$ & 0.063$^{***}$ & 0.141$^{***}$ & 0.121$^{***}$ & 0.106$^{***}$ & 0.110$^{***}$\\
& (0.020) & (0.017) & (0.020) & (0.015) & (0.018) & (0.013) & (0.016) & (0.011)\\  
IO (WIOD) & 0.055$^{***}$ & 0.017 & & & 0.063$^{***}$ & 0.057$^{***}$ & & \\
& (0.018) & (0.027) & & & (0.020) & (0.021) & & \\
IO (WIOD)*Labor & 0.003 & 0.002 & & & 0.008 & 0.011 & & \\
& (0.012) & (0.010) & & & (0.013) & (0.009) & & \\
IO (trans) & & & 0.031 & 0.026$^{*}$ & & & 0.098$^{***}$ & 0.040$^{***}$\\
& & & (0.030) & (0.015) & & & (0.023) & (0.011)\\ 
IO (trans)*Labor & & & 0.035$^{*}$ & 0.015$^{*}$ & & & 0.027$^{*}$ & 0.031$^{***}$ \\
& & & (0.018) & (0.008) & & & (0.014) & (0.007)\\
\hline \\[-1.8ex] 
\\[-2.5ex]Two way FE & No & Yes & No & Yes & No & Yes & No & Yes \\ 
\\[-2.5ex]Observations & 35,778 & 35,778 & 35,778 & 35,778 & 35,778 & 35,778 & 35,778 & 35,778 \\ 
R$^{2}$ & 0.014 & 0.255 & 0.014 & 0.256 & 0.027 & 0.243 & 0.034 & 0.245\\  
Adjusted R$^{2}$ & 0.014 & 0.243 & 0.014 & 0.244 & 0.027 & 0.232 & 0.034 & 0.233\\ 
\hline \\[-1.8ex] 
\end{tabular} 
\caption*{\textbf{Note:} Clustered (industry$_i$ and industry$_j$) standard errors in parentheses.  Significance codes: ***: p$<$0.01, **: p$<$0.05, *: p$<$0.1.}
\end{minipage}
\end{table}

\begin{table}[!htbp]
\begin{minipage}{\linewidth}
\centering
\fontsize{10}{12}\selectfont
\captionsetup{singlelinecheck=off, skip=0.ex}
\caption{OLS multivariate regressions with interaction effects excluding Budapest from the sample} 
\label{tab:si_interactions_budapestexcl} 
\begin{tabular}{@{\extracolsep{0.75pt}}lcccccccc} 
\\[-1.8ex]\hline 
\\[-1.8ex] & \multicolumn{4}{c}{Coagglomeration (EGK)} & \multicolumn{4}{c}{Coagglomeration (LC)} \\ 
\\[-2.5ex] & NUTS4 & NUTS4 & NUTS4 & NUTS4 & NUTS4 & NUTS4 & NUTS4 & NUTS4\\ 
\\[-2.5ex] & (1) & (2) & (3) & (4) & (5) & (6) & (7) & (8) \\ 
\hline \\[-1.8ex] 
Labor (SR) & 0.073$^{***}$ & 0.089$^{***}$ & 0.070$^{***}$ & 0.088$^{***}$ & 0.147$^{***}$ & 0.125$^{***}$ & 0.107$^{***}$ & 0.107$^{***}$\\
& (0.012) & (0.019) & (0.012) & (0.019) & (0.022) & (0.013) & (0.019) & (0.011) \\    
IO (WIOD) & 0.004 & 0.030$^{*}$ & & & 0.046$^{**}$ & 0.053$^{***}$ & & \\
& (0.008) & (0.017) & & & (0.021) & (0.016) & & \\ 
IO (WIOD)*Labor & 0.029$^{***}$ & 0.025$^{**}$  & & & 0.039$^{***}$ & 0.031$^{***}$ & & \\
& (0.010) & (0.011) & & & (0.013) & (0.009) & & \\ 
IO (trans) & & & 0.011 & 0.026$^{**}$ & & & 0.101$^{***}$ & 0.064$^{***}$ \\
& & & (0.010) & (0.011) & & & (0.016) & (0.010) \\
IO (trans)*Labor & & & 0.011 & 0.004 & & & 0.054$^{***}$ & 0.043$^{***}$ \\
& & & (0.010) & (0.012) & & & (0.010) & (0.007) \\ 
\hline \\[-1.8ex] 
\\[-2.5ex]Two way FE & No & Yes & No & Yes & No & Yes & No & Yes \\ 
\\[-2.5ex]Observations & 34,980 & 34,980 & 34,980 & 34,980 & 34,980 & 34,980 & 34,980 & 34,980 \\ 
R$^{2}$ & 0.007 & 0.029 & 0.006 & 0.029 & 0.029 & 0.262 & 0.042 & 0.267 \\ 
Adjusted R$^{2}$ & 0.007 & 0.015 & 0.006 & 0.014 & 0.029 & 0.251 & 0.042 & 0.256\\ 
\hline \\[-1.8ex] 
\end{tabular} 
\caption*{\textbf{Note:} Clustered (industry$_i$ and industry$_j$) standard errors in parentheses.  Significance codes: ***: p$<$0.01, **: p$<$0.05, *: p$<$0.1.}
\end{minipage}
\end{table}

\begin{table}[!htbp]
\begin{minipage}{\linewidth}
\centering
\fontsize{10}{12}\selectfont
\captionsetup{singlelinecheck=off, skip=0.ex}
\caption{OLS multivariate regressions with interaction effects (manufacturing only)} 
\label{tab:si_interactions_manuf} 
\begin{tabular}{@{\extracolsep{0.75pt}}lcccccccc} 
\\[-1.8ex]\hline 
\\[-1.8ex] & \multicolumn{4}{c}{Coagglomeration (EGK)} & \multicolumn{4}{c}{Coagglomeration (LC)} \\ 
\\[-2.5ex] & NUTS4 & NUTS4 & NUTS4 & NUTS4 & NUTS4 & NUTS4 & NUTS4 & NUTS4\\ 
\\[-2.5ex] & (1) & (2) & (3) & (4) & (5) & (6) & (7) & (8) \\ 
\hline \\[-1.8ex] 
Labor (SR) & 0.123$^{***}$ & 0.163$^{***}$ & 0.125$^{***}$ & 0.157$^{***}$ & 0.065$^{***}$ & 0.111$^{***}$ & 0.073$^{***}$ & 0.113$^{***}$\\
& (0.021) & (0.060) & (0.025) & (0.059) & (0.016) & (0.015) & (0.015) & (0.015)\\   
IO (WIOD) & 0.042$^{**}$  & 0.059 & & & 0.080$^{***}$ & 0.065$^{**}$ & & \\
& (0.019) & (0.057) & & & (0.018) & (0.028) & & \\
IO (WIOD)*Labor & -0.001 & 0.004 & & & 0.033$^{***}$ & 0.017$^{*}$  & & \\
& (0.011) & (0.014) & & & (0.010) & (0.009) & & \\
IO (trans) & & & 0.022 & 0.075 & & & 0.049$^{**}$  & 0.036$^{*}$\\
& & & (0.036) & (0.049) & & & (0.023) & (0.020)\\
IO (trans)*Labor & & & -0.004 & -0.021 & & & 0.040$^{***}$ & 0.024$^{**}$ \\
& & & (0.025) & (0.036) & & & (0.013) & (0.010) \\
\hline \\[-1.8ex] 
\\[-2.5ex]Two way FE & No & Yes & No & Yes & No & Yes & No & Yes \\ 
\\[-2.5ex]Observations & 4,536 & 4,536 & 4,536 & 4,536 & 4,536 & 4,536 & 4,536 & 4,536 \\ 
R$^{2}$ & 0.018 & 0.086 & 0.017 & 0.087 & 0.058 & 0.185 & 0.043 & 0.186 \\
Adjusted R$^{2}$ & 0.018 & 0.049 & 0.016 & 0.050 & 0.058 & 0.151 & 0.043 & 0.152 \\ 
\hline \\[-1.8ex] 
\end{tabular} 
\caption*{\textbf{Note:} Clustered (industry$_i$ and industry$_j$) standard errors in parentheses.  Significance codes: ***: p$<$0.01, **: p$<$0.05, *: p$<$0.1.}
\end{minipage}
\end{table}

\begin{table}[!htbp]
\begin{minipage}{\linewidth}
\centering
\fontsize{10}{12}\selectfont
\captionsetup{singlelinecheck=off, skip=0.ex}
\caption{OLS multivariate regressions with interaction effects (services only)} 
\label{tab:si_interactions_serv} 
\begin{tabular}{@{\extracolsep{0.75pt}}lcccccccc} 
\\[-1.8ex]\hline 
\\[-1.8ex] & \multicolumn{4}{c}{Coagglomeration (EGK)} & \multicolumn{4}{c}{Coagglomeration (LC)} \\ 
\\[-2.5ex] & NUTS4 & NUTS4 & NUTS4 & NUTS4 & NUTS4 & NUTS4 & NUTS4 & NUTS4\\ 
\\[-2.5ex] & (1) & (2) & (3) & (4) & (5) & (6) & (7) & (8) \\ 
\hline \\[-1.8ex] 
Labor (SR) & 0.107$^{***}$ & 0.023 & 0.066$^{***}$ & 0.020 & 0.309$^{***}$ & 0.122$^{***}$ & 0.186$^{***}$ & 0.087$^{***}$ \\
& (0.021) & (0.024) & (0.018) & (0.022) & (0.060) & (0.029) & (0.054) & (0.023)\\   
IO (WIOD) & 0.048 & 0.187$^{***}$ & & & -0.068$^{*}$  & 0.183$^{***}$ & & \\
& (0.030) & (0.064) & & & (0.040) & (0.058) & & \\
IO (WIOD)*Labor & -0.009 & 0.019 & & & -0.035 & 0.003 & & \\
& (0.019) & (0.018) & & & (0.031) & (0.021) & & \\
IO (trans) & & & 0.083$^{***}$ & -0.014 & & & 0.205$^{***}$ & 0.092$^{***}$ \\
& & & (0.030) & (0.023) & & & (0.045) & (0.024)\\
IO (trans)*Labor & & & 0.008 & 0.057$^{***}$ & & & 0.049$^{*}$ & 0.043$^{***}$ \\
& & & (0.014) & (0.017) & & & (0.027) & (0.016) \\
\hline \\[-1.8ex] 
\\[-2.5ex]Two way FE & No & Yes & No & Yes & No & Yes & No & Yes \\ 
\\[-2.5ex]Observations & 5,886 & 5,886 & 5,886 & 5,886 & 5,886 & 5,886 & 5,886 & 5,886 \\ 
R$^{2}$ & 0.006 & 0.128 & 0.009 & 0.126 & 0.062 & 0.505 & 0.098 & 0.508 \\ 
Adjusted R$^{2}$ & 0.006 & 0.095 & 0.009 & 0.092 & 0.062 & 0.486 & 0.097 & 0.489\\
\hline \\[-1.8ex] 
\end{tabular} 
\caption*{\textbf{Note:} Clustered (industry$_i$ and industry$_j$) standard errors in parentheses.  Significance codes: ***: p$<$0.01, **: p$<$0.05, *: p$<$0.1.}
\end{minipage}
\end{table}

\clearpage
\subsection*{S10 Additional visualisation of firm-to-firm connections}
\label{sec:si_firm_to_firm_ties}

\begin{figure}[!ht]
\includegraphics[width=0.95\textwidth]{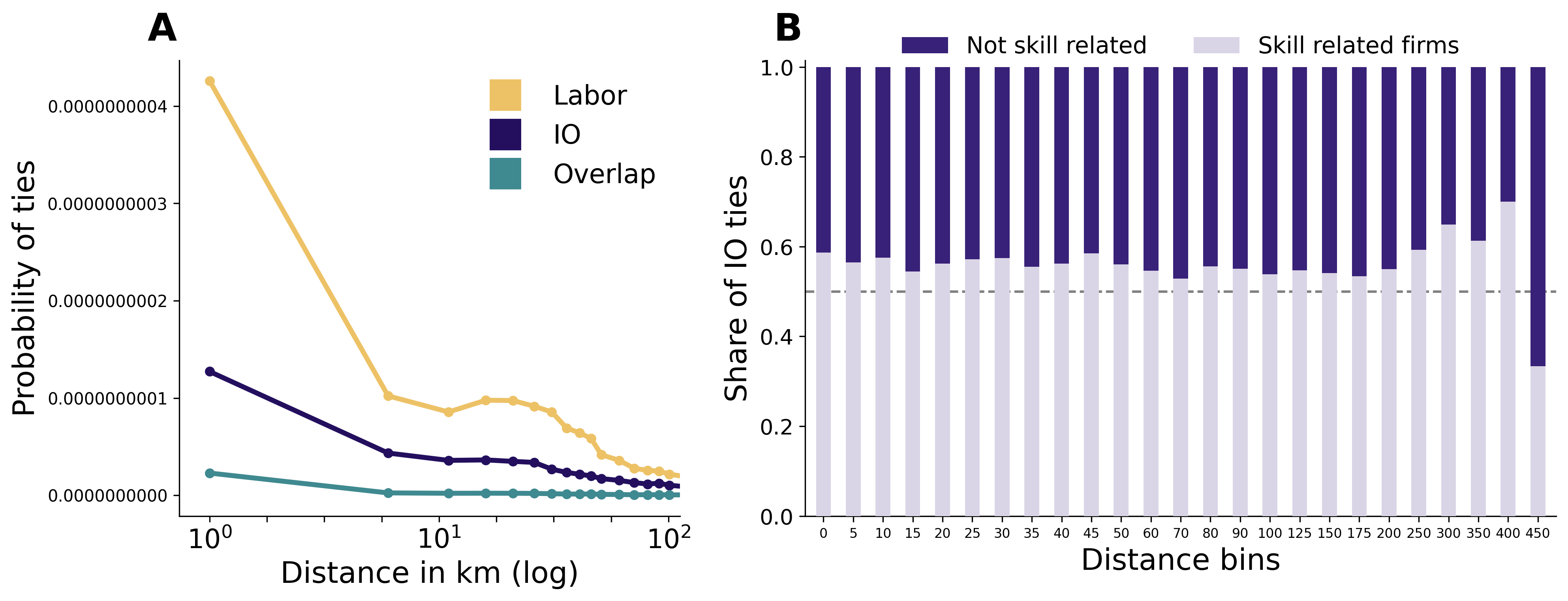} 
\caption{Alternative representations on the geography of firm-to-firm input-output (IO) and labor flow connections.
\textbf{(A)} Probability of labor flow, input-output connections and overlapping (labor and input-output) ties by distance.
\textbf{(B)} Share of input-output connections between firms in skill related (SR) and not skill related industries by distance bins. The figures are based on the sample of firms we used to construct our aggregate measures.}
\label{fig:io_distdistr}
\end{figure}

\end{document}